\documentclass[12pt]{article} 

\usepackage{etex}

\usepackage{amssymb,color}

\usepackage{graphics,graphicx,xcolor}

\newcommand{\nc}{\newcommand}
\nc{\la}{\lambda} \nc{\alf}{\alpha} \nc{\La}{\Lambda} \nc{\ze}{\zeta}
\nc{\tht}{\theta} \nc{\T}{\Theta} \nc{\be}{\beta}  \nc{\eps}{\epsilon} 
\nc{\ga}{\gamma}  \nc{\De}{\Delta}  \nc{\G}{\Gamma}  \nc{\vphi}{\varphi}
\nc{\de}{\delta} \nc{\si}{\sigma}  \nc{\ka}{\kappa}   \nc{\Si}{\Sigma} 
\nc{\om}{\omega}  \nc{\qq}{\quad\quad}                \nc{\Om}{\Omega}
\nc{\nf}{\infty}   \nc{\dl}{\mathop{\smash{\cal L}}}  \nc{\black}{\rule{3mm}{3mm}}
\nc{\ra}{\rightarrow}    \nc{\ol}{\overline}        \nc{\und}{\underline} 
\nc{\beq}{\begin{equation}}  \nc{\eeq}{\end{equation}}  \nc{\pt}{\partial}  
   \nc{\dst}{\displaystyle}  \nc{\na}{\nabla} 
\nc{\nnb}{\nonumber}    \nc{\bs}{\backslash}        \nc{\mb}{\mathbb}   
\nc{\sn}{{\rm sn}\,} \nc{\cn}{{\rm cn}\,}     \nc{\dn}{{\rm dn}\,} \nc{\nin}{\noindent}
\nc{\ti}{\tilde}   \nc{\wti}{\widetilde}   \nc{\h}{\hat}  \nc{\wh}{\widehat}
\nc{\tpsi}{\wti{\psi}}   \nc{\tphi}{\wti{\phi}}  \nc{\tH}{\wti{H}} \nc{\Ai}{{\rm Ai}}
\nc{\Pf}{P_{\phi}}

\nc{\rg}{\mathrm{g}}
\nc{\half}{\frac{1}{2}}

\newcounter{muni}
\newenvironment{remunerate}{\begin{list}{{\rm \arabic{muni}.}}
{\usecounter{muni}
\setlength{\leftmargin}{0pt}\setlength{\itemindent}{38pt}}}{\end{list}}

\nc{\brm}{\begin{remunerate}}   \nc{\erm}{\end{remunerate}}

\newtheorem{nth}{Proposition}  \newtheorem{nTh}{Theorem}   \newtheorem{nlem}{Lemma}  

\nc{\stg}{\mathop{\smash{*}}}
\nc{\st}{\mathop{\smash{\delta}}}
\nc{\barr}{\begin{array}}   \nc{\earr}{\end{array}}   \nc{\dg}{\dagger}
\nc{\mtvb}{\mathversion{bold}}   \nc{\mtvn}{\mathversion{normal}}  \nc{\F}{f_{\eps}}

\begin{document} 

\baselineskip=16.5pt

\begin{titlepage}

\hfill{July 31, 2014}

\vspace{1cm}
\centerline{\Large\bf Explicit metrics for a class of two-dimensional}
\vskip 0.5truecm
\centerline{\Large\bf {cubically} superintegrable systems}

\vskip 2.0truecm
\centerline{ \large\bf Galliano VALENT  \footnote{Sorbonne Universit\'es, UPMC Universit\'e Paris 06, UMR 7589, LPTHE, F-75005, 
Paris, France\\ \indent\indent CNRS, UMR 7589, LPTHE, F-75005, Paris, France.}$^{\, , \,\, 2}$
\indent\indent\indent Christian DUVAL \footnote{Aix-Marseille University, CNRS UMR 7332, CPT, 13288 Marseille, France.\\ 
\indent\indent Universit\'e de Toulon, CNRS UMR 7332, CPT, 83957 La Garde, France.}}

\vskip 1truecm
\centerline{ \large\bf Vsevolod SHEVCHISHIN  \footnote{Faculty of Mathematics\\ \indent \indent National Research University ``Higher School of Economics"\\ \indent \indent 7 Vavilova Str., Moscow, Russia.}}   

\vskip 2.5truecm

\begin{abstract} We obtain, in local coordinates, the explicit form of the two-dimensional, super\-integrable systems of Matveev and Shevchishin involving {linear} and cubic integrals. This enables us to deter\-mine for which values of the parameters these systems are indeed globally defined on~${\mb S}^2$.
\end{abstract}

\end{titlepage}

\newpage

%%%%%%%%%%%%%%%%%%%%%%%%%%%%%%%%%%%%%%%%%%%%%%%%%%%%%%%%%%%%%%%%%%%%%%%%%%%%%%%%%
%%%%%%%%%%%%%%%%%%%%%%%%%%%%%%%%%%%%%%%%%%%%%%%%%%%%%%%%%%%%%%%%%%%%%%%%%%%%%%%%%
\section{Introduction}
%%%%%%%%%%%%%%%%%%%%%%%%%%%%%%%%%%%%%%%%%%%%%%%%%%%%%%%%%%%%%%%%%%%%%%%%%%%%%%%%%
%%%%%%%%%%%%%%%%%%%%%%%%%%%%%%%%%%%%%%%%%%%%%%%%%%%%%%%%%%%%%%%%%%%%%%%%%%%%%%%%%

The study of superintegrable dynamical systems has received many 
important developments reviewed recently in \cite{mpw}. While integrable systems on the cotangent bundle $T^*M$ of a $n$-dimensional manifold, $M$, require a set of functionally independent observables 
$(H,Q_1,\ldots,Q_{n-1})$ which are all in involution for the Poisson bracket $\{\,\cdot\,,\,\cdot\,\}$, 
a super\-integrable system is made out of $\nu\geq n$ functionally independent observables 
\[
H,\qq Q_1,\qq Q_2, \quad\cdots\quad Q_{\nu-1},
\]
with the constraints 
\begin{equation}
\{H,Q_i\}=0,
\qquad
\hbox{for all\ }i=1,2,\ldots,\nu-1.
\label{HQ=0}
\end{equation}
The maximal value of $\nu$ is $2n-1$ since the system (\ref{HQ=0}) reads $dH(X_{Q_i})=0$, implying that the span of the Hamiltonian vector fields, $X_{Q_i}$, is, at each point of $T^*M$, a subspace of the annihilator of the $1$-form $dH$, the latter being of dimension $2n-1$. Let us observe that for 
two-dimensional manifolds, a superintegrable system is necessarily maximal since $\nu=3$.

{As is apparent from \cite{mpw}, the large amount of results for superintegrable models is restricted to 
{\em quadratically} superintegrable ones, which means that the integrals $Q_i$ are either linear or 
quadratic in the momenta, and the metrics on which these systems are defined are either flat or of 
constant curvature.}  
For manifolds of non constant curvature, Koenigs \cite{Ko} gave examples of quadratically super\-integrable   models. For some special values of the parameters the metrics happen to be defined on 
a manifold, $M$, which is never closed (compact without boundary).

In their quest for superintegrable systems defined on closed manifolds, Matveev and Shevchishin \cite{ms} have given a complete classification of all (local) Riemannian metrics on surfaces of revolution, namely 
\beq
\label{Ghhx}
G=\frac{dx^2+dy^2}{h_x^2},
\qq\qq
h=h(x),
\qq%\qq
h_x=\frac{dh}{dx},
\eeq
which have a superintegrable geodesic flow (whose Hamiltonian will henceforth be denoted by $H$), with  integrals $L=P_y$ and $S$ respectively linear and cubic in momenta, {opening the way to the new field 
of {\em cubically} superintegrable models.} Let us first recall their main results.

They proved that if the metric $G$ is not of constant curvature, 
then ${\cal I}^3(G)$, the linear span 
of the cubic integrals, has dimension $4$ with a natural basis $L^3,LH,S_1,S_2$, and with the 
following structure. The map ${\cal L}:S\to\{L,S\}$ defines a linear endomorphism of 
${\cal I}^3(g)$ and one of the following possibilities hold:

\goodbreak

\brm
\item[(i)] 
${\cal L}$ has purely real eigenvalues $\pm \mu$ for some real $\mu>0$, then $S_1,\,S_2$ 
are the corresponding eigenvectors.
\item[(ii)] 
${\cal L}$ has purely imaginary eigenvalues $\pm i\mu$ for some real $\mu>0$, then 
$S_1 \pm iS_2$ are the corresponding eigenvectors.
\item[(iii)] 
${\cal L}$ has the eigenvalue $\mu=0$ with one Jordan block of size $3$, in this case
\[
\{L,S_1\}=\frac{A_3}{2}\,L^3+A_1\,LH,
\qq\qq 
\{L,S_2\}=S_1,
\]
\erm
for some real constants $A_1$ and $A_3$. Superintegrability is then achieved provided the 
function $h$ be a solution of following non-linear first-order differential equations, namely
\beq
\label{odeglob}
\barr{crcl}
(i)& \qq  h_x(A_0\,h_x^2+\mu^2\,A_0\,h^2-A_1\,h+A_2) &=& \dst  A_3\,\frac{\sin(\mu\,x)}{\mu}
+A_4\,\cos(\mu\,x)\\[4mm] (ii)&\qq  h_x(A_0\,h_x^2-\mu^2\,A_0\,h^2-A_1\,h+A_2) &=& \dst  A_3\,\frac{\sinh(\mu\,x)}{\mu}+A_4\,\cosh(\mu\,x)\\[4mm] 
(iii)& \qq h_x(A_0\,h_x^2-A_1\,h+A_2) &=& A_3\,x+A_4
\earr
\eeq
and the explicit form of the cubic integrals was given in all three cases. For instance,  
when $\mu=1$ or $\mu=i$, their structure is 
\beq
\label{F12}
S_{1,2}=e^{\pm\mu y}\left(
 a_0(x)\,P_x^3+a_1(x)\,P_x^2\,P_y+a_2(x)\,P_x\,P_y^2+a_3(x)\,P_y^3
 \right),%\\
\eeq
where the $a_i(x)$ are explicitly expressed in terms of $h$ and its derivatives; see \cite{ms}. 

For $A_0=0$ these equations are easily integrated and one obtains the Koenigs metrics~\cite{Ko}, while the cubic integrals have the reducible structure $S_{1,2}=P_y\,Q_{1,2}$ where the quadratic integrals $Q_{1,2}$ are precisely those obtained by Koenigs. 

Furthermore it was proved that in the case $(ii)$, under the conditions
\beq\label{condMS}
\mu>0,
\qq\qq A_0>0,
 \qq\qq 
\mu\,A_4>|A_3|,
\eeq
the metric and the cubic integrals are real-analytic and globally defined on ${\mb S}^2$.

The aim of this article is on the one hand to integrate explicitly the three differential equations in (\ref{odeglob}) 
and, on the other hand, to determine, by a systematic case study, all special cases which lead to superintegrable models \textit{globally} defined on simply-connected, \textit{closed}, Riemann surfaces.

\goodbreak

In Section \ref{trigoSection} we analyze the trigonometric case (real eigenvalues), integrating explicitly the differential equation (\ref{odeglob},$i$) to get an explicit local form for the metric and the cubic integrals. The 
global questions are then discussed, and we show that there is no closed manifold, $M$, on which the superintegrable model under consideration can be defined.

In Section \ref{hyperSection} we investigate the hyperbolic case (purely imaginary eigenvalues). Here too, the integration of the 
differential equation (\ref{odeglob},$ii$) provides an explicit form for both the metric and the cubic integrals. 

The previous results allows the determination of all superintegrable systems globally defined on ${\mb S}^2$, and these are proved in Theorem \ref{Thm1} and Theorem \ref{Thm2}, namely

\begin{nTh} 
The metric 
\[
G=\rho^2\,\frac{dv^2}{D}+\frac{4D}{P}\,d\phi^2,
\qq\qq%\qq 
v\in(a,1),
\qq\qq
\phi\in{\mb S}^1,
\]
with
\beq
D=(v-a)(1-v^2),
\qq%\qq 
P=(v^2-2av+1)^2,
\qq 
-\rho=1+4\frac{(v-a)D}{P},
\eeq
is globally defined on ${\mb S}^2$, as well as the Hamiltonian
\[
H=\half\,G^{ij}P_iP_j=\half\left(\Pi^2+\frac{P}{4D}\Pf^2\right),
\qq 
\Pi=\frac{\sqrt{D}}{\rho}\,P_v,
\]
iff $a\in(-1,+1)$. The two cubic integrals $S_1$ and $S_2$, also globally defined on ${\mb S}^2$, read
\beq 
\label{Sintro1}
S_1=\cos\phi\,{\cal A}+\sin\phi\,{\cal B}, 
\qq%\qq 
S_2=-\sin\phi\,{\cal A}+\cos\phi\,{\cal B},
\eeq
where
\beq
\label{Sintro2}
{\cal A}=\Pi^3-f\,f''\,\Pi\,\Pf^2,
\qq 
{\cal B}=f'\,\Pi^2\,\Pf-f\,(1+f'\,f'')\,\Pf^3,
\qq \qq
f=\sqrt{D}.
\eeq
\end{nTh}
%and
\begin{nTh} 
The metric
\beq
G=\rho^2\,\frac{dx^2}{D}+\frac{4D}{P}\,d\phi^2,
\qq 
\rho=\frac{Q}{P},
\qq%\qq 
x\in(-1,+1),
\qq 
\phi\in{\mb S}^1,
\eeq
with
\beq
\left\{
\barr{ll}
D=(x+m)(1-x^2),
&  \\[4mm]
P=\Big(L_+\,(1-x^2)+2(m+x)\Big)\Big(L_-\,(1-x^2)+2(m+x)\Big), 
& \qq L_{\pm}=l\pm\sqrt{l^2-1},\\[4mm] 
Q=3x^4+4mx^3-6x^2-12mx-4m^2-1,
&%\dst   
\earr
\right.
\eeq
is globally defined on ${\mb S}^2$, as well as the Hamiltonian
\[
H=\half\,G^{ij}P_iP_j=\half\left(\Pi^2+\frac{P}{4D}\Pf^2\right),
\qq 
\Pi=\frac{\sqrt{D}}{\rho}\,P_x,
\]
iff $m>1$, and $l>-1$.
The two cubic integrals $S_1$ and $S_2$, still given by the formulas (\ref{Sintro1}) and (\ref{Sintro2}), are 
also globally defined on ${\mb S}^2$.
\end{nTh}

{
In Subsection \ref{newSubsection} we compare of our results with those
of Matveev and Shevchishin \cite{ms}. In particular, for a convenience of the reader, we provide the transition formulas between the coordinates and
functions used in \cite{ms} and the coordinates and function used in the present paper.
}

In Section \ref{affineSection} we analyze the affine case (zero eigenvalue). As in the trigonometric case, the 
system is never defined on closed manifolds but we determine in which cases it is  
globally defined either on ${\mb R}^2$ or on ${\mb H}^2$.

In Section \ref{conclusionSection} we draw some conclusions and present some possibly interesting strategy for future developments.

\bigskip
\textbf{Acknowledgements:} We wish to warmly thank V. Matveev, and J.-P. Michel for their interest in this work, and for enlightening discussions.
%most valuable suggestions.

%%%%%%%%%%%%%%%%%%%%%%%%%%%%%%%%%%%%%%%%%%%%%%%%%%%%%%%%%%%%%%%%%%%%%%%%%%%%%%%%%
%%%%%%%%%%%%%%%%%%%%%%%%%%%%%%%%%%%%%%%%%%%%%%%%%%%%%%%%%%%%%%%%%%%%%%%%%%%%%%%%%
\section{The trigonometric case}\label{trigoSection}
%%%%%%%%%%%%%%%%%%%%%%%%%%%%%%%%%%%%%%%%%%%%%%%%%%%%%%%%%%%%%%%%%%%%%%%%%%%%%%%%%
%%%%%%%%%%%%%%%%%%%%%%%%%%%%%%%%%%%%%%%%%%%%%%%%%%%%%%%%%%%%%%%%%%%%%%%%%%%%%%%%%

%%%%%%%%%%%%%%%%%%%%%%%%%%%%%%%%%%%%%%%%%%%%%%%%%%%%%%%%%%%%%%%%%%%%%%%%%%%%%%%%%
\subsection{The explicit form of the metric}
%%%%%%%%%%%%%%%%%%%%%%%%%%%%%%%%%%%%%%%%%%%%%%%%%%%%%%%%%%%%%%%%%%%%%%%%%%%%%%%%%

The ode (\ref{odeglob},$i$) obtained in \cite{ms} is:
\[
h_x\Big(A_0\,h_x^2+\mu^2\,A_0\,h^2-A_1\,h+A_2\Big)=A_3\frac{\sin(\mu\,x)}{\mu}+A_4\,\cos(\mu\,x).
%\qq h_x=D_x h
\]
For the Koenigs metrics $A_0=0$; we thus must consider here a non-vanishing $A_0$ which can be scaled to 1. 
By a scaling of $x$ we can also set $\mu=1$. By a translation of $x$ and a scaling of $h$ the right-hand side  becomes $\la\,\sin x$, with $\la$ a free real parameter. By a translation of $h$, we can set $A_1=0$ and $A_2=a$. We hence have to solve
\beq\label{ode}
h_x(h_x^2+h^2+a)=\la\,\sin x,
\qq\qq 
a\in{\mb R},
\qq 
\la\in{\mb R}\bs\{0\}.
\eeq
Let us regard now $u=h_x$ as a function of the variable $h$ and define
\beq
U=u(u^2+h^2+a)
\qq\mbox{with}\qq 
\frac{d^2U}{dx^2}+U=0.
\eeq
This last relation, when expressed in terms of the variable $h$ becomes then
\beq
\frac{d}{dh}\left(u\,\frac{dU}{dh}\right)+u^2+h^2+a=0,
\qq\qq 
a\in{\mb R},
\eeq
and can be integrated, yielding
% up to
\beq\label{dU}
4hu\,\frac{dU}{dh}=c+(u^2+h^2+a)(3u^2-h^2-a).
\eeq
Since $U=\la\,\sin x$ we have also a first order equation
\beq
U'^2=\la^2-U^2
\qq
\Rightarrow
\qq 
\left(4hu\,\frac{dU}{dh}\right)^2=16h^2\,(\la^2-U^2),
\eeq
and upon using (\ref{dU}) we obtain a quartic equation for $u$:
\beq
\Big[c+(u^2+h^2+a)(3\,u^2-h^2-a)\Big]^2=16\,h^2\Big[\la^2-u^2(u^2+h^2+a)^2\Big].
\eeq
If we define $v=u^2+h^2$, this equation remains a quartic in $v$ but happens to be linear in~$h^2$. Solving for $h^2$ in terms of the variable $v$, we find
\beq
\label{h2i}
v=u^2+h^2,
\qq%\qq
h^2=\frac{D'^2}{8D},
\qq%\qq  
D(v)=(v+a)(v^2-a^2+c)+2\la^2.
\eeq
At this stage, it turns out to be convenient to define 
\beq
f=\sqrt{D}=\sqrt{(v+a)(v^2-a^2+c)+2\la^2} 
\qq
\hbox{and}
\qq  
g=2v-f'^2
\eeq
where $f'=df/dv$. This allows, once the old coordinates $(x,y)$ have been expressed in terms of the new ones, $(v,y)$, to get eventually the explicit form of the metric
\beq
\label{metcasi}
\half G=\frac{1}{2h_x^2}(dx^2+dy^2)
=\left(\frac{f''}{g}\right)^2dv^2+\frac{dy^2}{g}
\eeq
which gives the Hamiltonian 
\beq
H\equiv G^{ij}P_iP_j=\half\left(\Pi^2+g\,P_y^2\right),
\qq\qq 
\Pi=\frac{g}{f''}\,P_v.
\eeq

%%%%%%%%%%%%%%%%%%%%%%%%%%%%%%%%%%%%%%%%%%%%%%%%%%%%%%%%%%%%%%%%%%%%%%%%%%%%%%%%%
\subsection{The cubic integrals}
%%%%%%%%%%%%%%%%%%%%%%%%%%%%%%%%%%%%%%%%%%%%%%%%%%%%%%%%%%%%%%%%%%%%%%%%%%%%%%%%%

They were given in (\ref{F12}), as borrowed from \cite{ms}, and  become in our new coordinates with a slight change of notation 
\beq
S_{\pm}=e^{\pm y}\Big(\Pi^3\mp f'\,\Pi^2\,P_y+f\,f''\,\Pi\,P_y^2\pm f(1-f'f'')P_y^3\Big).
\eeq
However due to the relation $\,dH\wedge dP_y\wedge dS_+\wedge dS_-=0$, the four observables involved are not functionally independent. Indeed, we have
\beq
S_+\,S_-= 8H^3+8a\,H^2\,P_y^2+2c\,H\,P_y^4-2\la^2\,P_y^6,
\eeq
so that we may consider two different super\-integrable systems
\begin{equation}
\label{I+I-}
{\cal I}_+=\,(H,\,P_y,\,S_+)
\qq
\hbox{and}
\qq 
{\cal I}_-=\,(H,\,P_y,\,S_-).
\end{equation}

%Let us prove

\begin{nth} 
The observables $\,S_+$ and $S_-$ are integrals and the set $(H,\,P_y,\,S_+,\,S_-)$ generates 
a Poisson algebra.
\end{nth}
\nin{\bf Proof:} 
The Poisson brackets are given by
\beq
\label{PBHSpm}
\{H,S_{\pm}\}=e^{\pm y}\,\frac{g}{f''}\,\Pi\,P_y^2(\Pi \mp f'\,P_y)\,
\Big(f\,f'''-3(1-f'\,f'')\Big).
\eeq
Quite remarkably, the ode  
\beq
\label{odespecialei}
f\,f'''-3(1-f'\,f'')=0
\eeq
does linearize upon the substitution $f=\sqrt{D}$ since we have
\beq
2\Big(f\,f'''-3(1-f'\,f'')\Big)=D'''-6=0,
\eeq
which gives for $D$ the most general monic polynomial of third degree 
\beq
\label{Di}
D(v)=v^3-s_1\,v^2+s_2\,v-s_3,
\eeq 
whose coefficients are expressed in terms of the symmetric functions of the roots. As a matter of fact, the function $D$ already obtained in (\ref{h2i}) displays exactly 
$3$ parameters $a,c,\lambda$. Equations (\ref{PBHSpm}) and (\ref{odespecialei}) insure then conservation of both cubic integrals $S_+$ and~$S_-$.

The Poisson algebra structure follows from the following relations, viz.,
\beq
\barr{rcl}
\{S_+,S_-\} & = & \dst -16a\,H^2\,P_y-8c\,H\,P_y^3+12\la^2\,P_y^5,\\[4mm] 
S_+\,S_- & = & \dst 8H^3+8a\,H^2\,P_y^2+2c\,H\,P_y^4-2\la^2\,P_y^6;
\earr
\eeq
it is generated by $4$ observables in this case. 
$\quad\Box$ 

%%%%%%%%%%%%%%%%%%%%%%%%%%%%%%%%%%%%%%%%%%%%%%%%%%%%%%%%%%%%%%%%%%%%%%%%%%%%%%%%%
\subsection{Transformation of the metric and its curvature}
%%%%%%%%%%%%%%%%%%%%%%%%%%%%%%%%%%%%%%%%%%%%%%%%%%%%%%%%%%%%%%%%%%%%%%%%%%%%%%%%%

Taking for $D$ the expression (\ref{Di}), let us define the following quartic polynomials $P$ and~$Q$, namely
\beq
P=8v\,D-D'^2,
\qq\qq 
Q=2\,D\,D''-D'^2=P+4(v-s_1)D,
\qq 
Q'=12\,D,
\eeq
enabling us to write the metric (\ref{metcasi}) in the form
\beq\label{Gcasi}
\half G=\rho^2\,\frac{dv^2}{D}+\frac{4D}{P}\,dy^2,
\qq\qq 
\rho\equiv \frac{Q}{P}=1+(v-s_1)\frac{4D}{P},
\eeq
the scalar curvature being given by
\beq
\label{Ri}
R_G=\frac{1}{4Q^3}\Big(2PQ\,W'-(QP'+2PQ')\,W\Big),
\qq 
W\equiv DP'-PD'=8D^2-QD'.
\eeq
One should bear in mind the following restrictions:
\brm
\item The relation $v=u^2+h^2$ requires $v>0$.
\item For $h$ to be real we must have $D>0.$ 
\item For the metric $G$ to be Riemannian we need $P>0$.
\erm

%%%%%%%%%%%%%%%%%%%%%%%%%%%%%%%%%%%%%%%%%%%%%%%%%%%%%%%%%%%%%%%%%%%%%%%%%%%%%%%%%
\subsection{Global properties}
%%%%%%%%%%%%%%%%%%%%%%%%%%%%%%%%%%%%%%%%%%%%%%%%%%%%%%%%%%%%%%%%%%%%%%%%%%%%%%%%%

To study the global geometry of these superintegrable models, we will be using  techniques which have proved quite successful in \cite{Va1} and \cite{Va2} for integrable models with either a cubic or 
a quartic integral. 

As emphasized in the Introduction, we will from now on confine considerations to the case of simply connected Riemann surfaces, which, by the Riemann uniformization theorem~\cite{Gu}, are conformally related to spaces of constant curvature ${\mb S}^2,{\mb R}^2,{\mb H}^2$.

One has first to determine, from the above positivity conditions, the open interval $I\subset{\mb R}$ admissible for the variable $v$. The end-points are singular points for the metric and 
the possibility of a manifold structure is related to the behavior of the metric at these end-points. 
Either they are true singularities (for instance if the scalar curvature is divergent at these points) 
or they are apparent singularities (also called coordinate singularities) due to a bad choice of the 
coordinates as, for instance,
\beq
G=dr^2+r^2\,d\phi^2,
\qq\qq 
r\in(0,+\nf),
\qq 
\phi\in{\mb S}^1,
\eeq
for which $r=0$ is an apparent singularity which can be wiped out, using back Cartesian coordinates. 

We will detect true singularities from the scalar curvature:
\begin{nlem}
\label{Lemma1}
 Let us consider the interval $I=(a,b)$, allowed for $v$, i.e., such that $D(v)>0$ and $P(v)>0$ for all $v\in\,I$. Suppose that $Q$ has a simple real zero $v_*\,\in\,I$; then $v=v_*$ is a curvature singularity precluding any manifold structure associated with the metric. 
\end{nlem}

\nin{\bf Proof:} The relation (\ref{Ri}) entails that
\beq
\lim_{v\to v_*}\,Q^3(v)\,R_G(v)=-4\,P(v_*)\,D^2(v_*)\,Q'(v_*)
\eeq
and the right-hand side of this equation does not vanish. The existence of such a 
curvature singularity for $v_*\in I$ rules out the possibility of a manifold structure. 
$\quad\Box$

We will detect non-closedness by 
\begin{nlem} 
\label{Lemma2}
If the variable $v$ takes its values in some interval $I=(a,b)$ and if one of the 
end-points is a zero of $P$ (and not of $Q$), then the manifold having infinite measure, it cannot be closed. 
\end{nlem}
\nin{\bf Proof:} Let the allowed interval for $v$ be $I=(a,b)$. The measure of the manifold is
\beq
\label{vol}
\mu_G=4\int_a^b{\!\frac{Q(v)}{P^{3/2}(v)}\,dv}\int{\!dy}.
\eeq
Now, if $P$ has a zero at one end-point where $Q$ does not vanish, the 
this integral 
%(\ref{vol}) 
will be divergent. $\ \Box$

\goodbreak

Let us turn ourselves to the analysis of this first case ($i$). Given any polynomial $P$ we will use 
the notation $\De(P)$ for its discriminant. The discussion will be organized according to 
the sign of $\De(D)$. Let us begin with:

\begin{nth}
\label{Prop2}
If $\De(D)=0$ the super\-integrable systems ${\cal I}_+$ and ${\cal I}_-$ given by (\ref{I+I-}) are either 
trivial or are not defined on a closed manifold.
\end{nth}
\nin{\bf Proof:} 
If $\De(D)=0$, we may have first $D=(v-v_0)^3$. The scalar curvature, easily 
computed using (\ref{Ri}), is a constant. The following theorem, due to Thompson \cite{Th}, states that \textit{for Riemannian spaces of constant curvature, namely
$\ {\mb S}^n,\ {\mb R}^n,\ {\mb H}^n\ $ with $n\geq 2$, every (symmetric) Killing-St\"ackel tensor 
of {\em any degree} is fully reducible to symmetrized tensor products of the Killing vectors.}
This implies that the cubic integrals are reducible, leaving us with the trivial integrable 
system $(H,\,P_y)$.

For $\De(D)=0$ we may also have $D=(v-v_0)(v-v_1)^2$ with $v_0\neq v_1$, which yields
\beq
\left\{
\barr{l} 
P(v)=-(v-v_1)^2\,p(v),
\qq 
p(v)=v^2-2(2v_0+3v_1)v+(2v_0+v_1)^2,\\[4mm]
\dst  
Q=3(v-v_1)^3(v-v_*),
\qq
v_*=v_0+\frac{v_0-v_1}{3}.
\earr
\right.
\eeq
Let us first observe that for the metric
\beq
\half G=\frac{9(v-v_*)^2}{p(v)^2}\,\frac{dv^2}{(v-v_0)}+\frac{4(v-v_0)}{(-p(v))}\,dy^2\eeq
to be Riemannian we must have $v>v_0$ and $p(v)<0$. If the roots $w_{\pm}$ of $p$ are ordered 
as $w_-<w_+$, positivity of the metric is achieved iff $v\in\, I=(v_0,+\nf)\cap(w_-, w_+)$, the 
upper bound of $I$ being $w_+$. Since $P(w_+)=0$ and $Q(w_+)\neq 0$, the expected manifold cannot be closed by Lemma \ref{Lemma2}. $\quad\Box$ 

%Let us proceed to:

\begin{nth} 
If $\De(D)<0\,$ the super\-integrable systems $\,{\cal I}_+$ and $\,{\cal I}_-$ given by (\ref{I+I-}) 
are never globally defined on a closed manifold.
\end{nth}

\nin{\bf Proof:} 
If $\De(D)<0\,$ the polynomial $D$ has only a simple real zero. Using new parameters $(a,\,b)$ we can write 
\[
D=(v-v_0)\Big((v-a)^2+b^2\Big),
\qq  
v\in\,(v_0,+\nf),
\qq 
a\in{\mb R},
\quad 
b\in{\mb R}\bs\{0\},
\]
with
\[\De(D)=-4b^2\Big((v_0-a)^2+b^2\Big)^2\]
and, for $P$ and $Q$, 
\beq
\De(P)=16384\,a^2\Big((v_0+a)^2+b^2\Big)^2\,\De(D),
\qq 
\De(Q)=27648\,b^2\Big((v_0-a)^2+b^2\Big)^2\,\De(D).
\eeq
We must exclude $a=0$ since $P(v)=-(v^2-2v_0v-b^2)^2$ is negative. Hence, the previous discriminants are strictly negative, implying that both polynomials $P$ and $Q$ have two simple real zeroes. 

The relation $Q'=12\,D$ shows that $Q$ is strictly increasing from $Q(v_0)=-[(v_0-a)^2+b^2]$ to $Q(+\nf)=+\nf$, hence there exists a simple zero $v_*$ of $Q$ such that $v_*>v_0$ while the other one 
lies to the left of $v_0$ because $Q(-\nf)=+\nf$.

The polynomial $P$ retains the form
\[P(v)=-\Big(v^2-2(v_0+2a)v-a^2-b^2-2av_0\Big)^2+16a\Big((v_0+a)^2+b^2\Big)v\]
showing that for $a<0$ it is never positive as it should; so, we are left with the case $a>0$. 
From the relations 
\[
P(v)=Q(v)+4(v_0+2a-v)\,D(v),
\qq\qq 
P'(v)=8\,D(v)+4(2a+v_0-v)\,D'(v),
\]
we see that $P(v_0)$ is strictly negative and that $P'(v)$ is positive from $v=v_0$ to $v=v_0+2a$. 
Thus $P$ increases to its first zero $v=w_-<v_*$ (since $P(v_*)=4(2a+v_0-v_*)\,D(v_*)>0$), 
is equal to $Q$ for $v=v_0+2a>v_*$, then vanishes at its second 
zero $w_+$ such that $w_+>v_0+2a$ and, at last, decreases to $-\nf$. 
Therefore, we end up with the ordering
\[v_0<w_-<v_*<v_0+2a<w_+.\]
So, $D>0$ and $P>0$ iff $v\in(w_-,w_+)$, and within this interval $Q$ has a simple zero 
for $v=v_*$; hence, by Lemma \ref{Lemma1}, there is no underlying manifold structure. 
$\quad\Box$

Let us conclude this section with
\begin{nth} 
If $\De(D)>0\,$ the super\-integrable systems $\,{\cal I}_+$ and $\,{\cal I}_-$ given by (\ref{I+I-}) 
are never globally defined on a closed manifold.
\end{nth}

\nin{\bf Proof:} Let us order the roots of $D$ according to $0\leq v_0<v_1<v_2$, so that
% we have
\[
D(v)=(v-v_0)(v-v_1)(v-v_2)=v^3-s_1\,v^2+s_2\,v-s_3,
\]
and $D>0$ for $v\in (v_0,v_1)\cup (v_2,+\nf)$. We need to determine now the positivity interval for $P$. 
Since 
\[
\De(P)=4096\,\si^2\De(D)>0,
\qq\qq 
\si=(v_0+v_1)(v_1+v_2)(v_2+v_0)>0,
\]
there will be either four real simple roots or no real root for $P$. The latter is excluded since $P=8vD-(D')^2$ is negative at the zeroes of $D$, and positive at those of $D'$. 
Also, notice that $\De(Q)=-6912\,\De^2(D)<0$ implies that $Q$ has two simple real roots 
and one of them is $v_*>v_2$. This is so because $Q(v)=P(v)+4(v-s_1)D(v)$, which shows 
that $Q(v_2)=P(v_2)=-(v_0-v_2)^2(v_1-v_2)^2<0$; but $Q'=12D$ entails that, for 
positive $D$, the function $Q$ is increasing with $Q(+\nf)=+\nf$. Hence $v=v_*$ is a simple zero of $Q$, forbidding any manifold structure by Lemma \ref{Lemma1}.

The zeroes of $P$ may appear only when $D>0$. Let us consider $v\in (v_0,v_1)$. Observing that $P(v_0)=-(v_0-v_1)^2(v_0-v_2)^2$ and $P(v_1)=-(v_1-v_0)^2(v_1-v_2)^2$ are negative 
and that there does exist $v=v_-\in (v_0,v_1)$ for which $D'(v_-)=0$, we get $P(v_-)>0$ which implies $v_0<w_0<v_-<w_1<v_1$, where $(w_0,w_1)$ is the first pair of simple zeroes of $P$. Positivity of both $D$ and $P$ is therefore obtained for $v\in(w_0,w_1)$. The function $Q$ remains strictly negative for $v\in\,[v_0,v_1]$,  and Lemma \ref{Lemma2} help us conclude that the supposed manifold cannot be closed.

The remaining two zeroes of $P$ denoted by $w_2<w_3$ must lie in $(v_2,+\nf)$. Since $Q(v_2)=-(v_2-v_0)^2(v_2-v_1)^2<0$ and then it increases to $Q(+\nf)=+\nf$ it will have 
a simple zero $v=v_*>v_2$, and at this point $P(v_*)=4(s_1-v_*)D(v_*)$. Let us discuss:
\brm
\item 
If $v_*<s_1$, we have $P(v_*)>0$, and since $P(+\nf)=-\nf$ we get $v_2<w_2<v_*<w_3$.  
The positivity of $D$ and $P$ requires $v\in (w_2,w_3)$, and there is no manifold structure since the curvature $R_G$ is singular at $v=v_*$.
\item 
If $v_*\geq s_1$, we have $P(v_*)<0$ hence $v_2<v_*<w_2<w_3$, and the positivity of $D$ and $P$ 
requires $v\in (w_2,w_3)$. Since $Q(w_3)>0$ the supposed manifold cannot be closed by Lemma \ref{Lemma2}. $\quad\Box$
\erm

We conclude this section by observing that the trigonometric case never leads to superintegrable systems defined on a closed manifold.

%%%%%%%%%%%%%%%%%%%%%%%%%%%%%%%%%%%%%%%%%%%%%%%%%%%%%%%%%%%%%%%%%%%%%%%%%%%%%%%%%
%%%%%%%%%%%%%%%%%%%%%%%%%%%%%%%%%%%%%%%%%%%%%%%%%%%%%%%%%%%%%%%%%%%%%%%%%%%%%%%%%
\section{The hyperbolic case}\label{hyperSection}
%%%%%%%%%%%%%%%%%%%%%%%%%%%%%%%%%%%%%%%%%%%%%%%%%%%%%%%%%%%%%%%%%%%%%%%%%%%%%%%%%
%%%%%%%%%%%%%%%%%%%%%%%%%%%%%%%%%%%%%%%%%%%%%%%%%%%%%%%%%%%%%%%%%%%%%%%%%%%%%%%%%

%%%%%%%%%%%%%%%%%%%%%%%%%%%%%%%%%%%%%%%%%%%%%%%%%%%%%%%%%%%%%%%%%%%%%%%%%%%%%%%%%
\subsection{The explicit form of the metric}
%%%%%%%%%%%%%%%%%%%%%%%%%%%%%%%%%%%%%%%%%%%%%%%%%%%%%%%%%%%%%%%%%%%%%%%%%%%%%%%%%

The ode (\ref{odeglob},$ii$) obtained in \cite{ms} is
\beq
h_x(A_0\,h_x^2-\mu^2\,A_0\,h^2-A_1\,h+A_2)=A_3\frac{\sinh(\mu\,x)}{\mu}+A_4\,\cosh(\mu\,x).
\eeq
Again, we may put $A_0=1,\mu=1,A_1=0,A_2=-a$, but, this time, the right-hand side of the previous equation leads to three different cases we will describe according to
\beq
\label{trigoEq}
h_x(h_x^2-h^2-a)=\frac{\la}{2}\,(e^{x}+\eps\,e^{-x}),
\qq\qq \eps=0,\pm1
\eeq
where $\la$ is a free parameter. 

\goodbreak

Let us point out that for $\eps=0$ the changes $x \to -x$ 
and $\la \to -\la$ show that there is no need to consider $e^{-x}$ in the right-hand side of (\ref{trigoEq}). 

With the definitions
\[
u=h_x,
\qq\qq 
U=u(u^2-h^2-a),
\qq\qq
a\in{\mb R},
\]
we get similarly
\[
U''-U=0
\qq
\Rightarrow
\qq
\frac{d}{dh}\left(u\,\frac{dU}{dh}\right)-(u^2-h^2-a)=0,
\]
which can be integrated to yield
\beq 
\label{equtilei}
4hu\,\frac{dU}{dh}=c+(u^2-h^2-a)(3u^2+h^2+a),
\qq\qq   c\in{\mb R}.
\eeq
Since $\dst U=\frac{\la}{2}\,(e^{x}+\eps\,e^{-x})$ we also have the first order ode:
\beq
U'^2=U^2-\eps\,\la^2
\qq
\Rightarrow
\qq 
\left(4hu\,\frac{dU}{dh}\right)^2=16\,h^2\,(U^2-\eps\la^2),
\eeq
which, upon use of (\ref{equtilei}), leaves  us with a quartic equation in the variable $u$. 
Positing $v=h^2-u^2$, we still have a quartic in $v$ but the $h^2$ dependence is 
merely linear and we can solve for $h^2$ in terms of the variable $v$, namely 
\beq
v=u^2-h^2,
\qq\qq 
h^2=\frac{D'^2}{8D},
\qq\qq 
D(v)=(a-v)(v^2-a^2+c)-2\eps\la^2,
\eeq
giving a result surprisingly similar to the case ($i$), except that $v$ needs not be positive. 
Upon defining
\beq
\label{fgii}
f=\sqrt{D}=\sqrt{(a-v)(v^2-a^2+c)-2\eps\la^2}
\qq
\hbox{and}
\qq  
g=f'^2+2v,
\eeq
we obtain the metric in the new coordinates $(v,y)$ in the form
\beq\label{metcasii}
\half G=\frac{1}{2h_x^2}(dx^2+dy^2)=\left(\frac{f''}{g}\right)^2\,dv^2+\frac{dy^2}{g},
\eeq
together with the Hamiltonian
\beq
H\equiv G^{ij}\,P_i\,P_j=\half\left(\Pi^2+g\,P_y^2\right),
\qq\qq 
\Pi=\frac{g}{f''}\,P_v.
\eeq

%%%%%%%%%%%%%%%%%%%%%%%%%%%%%%%%%%%%%%%%%%%%%%%%%%%%%%%%%%%%%%%%%%%%%%%%%%%%%%%%%
\subsection{The cubic integrals}
%%%%%%%%%%%%%%%%%%%%%%%%%%%%%%%%%%%%%%%%%%%%%%%%%%%%%%%%%%%%%%%%%%%%%%%%%%%%%%%%%

They were given in (\ref{F12}) 
%see \cite{ms} 
and read in our coordinates
\beq\label{Scasii1} 
S_1=\cos y\,{\cal A}+\sin y\,{\cal B},
\qq\qq 
S_2=-\sin y\,{\cal A}+\cos y\,{\cal B},
\eeq
where
\beq
\label{Scasii2}
{\cal A}=\Pi^3-f\,f''\,\Pi\,P_y^2,
\qq 
{\cal B}=f'\,\Pi^2\,P_y-f\,(1+f'\,f'')\,P_y^3.
\eeq

%Let us prove:

\begin{nth} 
The observables $S_1$ and $S_2$ are integrals of the geodesic flow.
\end{nth}

\nin{\bf Proof:} 
Let us define the complex object
\beq
{\cal S}=S_1+iS_2=e^{-iy}({\cal A}+i{\cal B}).
\eeq
The Poisson bracket with the Hamiltonian reads
\beq 
\label{PBHS}
\{H,{\cal S}\}=-e^{-iy}\frac{g}{f''}\Pi\,P_y^2(\Pi+if'\,P_y)\Big(f\,f'''+3(1+f'\,f'')\Big).
\eeq
Again, the transformation $f=\sqrt{D}$ leads to the following linearization:
\beq
\label{Dii}
2\Big(f\,f'''+3(1+f'\,f'')\Big)=D'''+6=0
\qq\Longrightarrow\qq 
D=-(v^3-s_1\,v^2+s_2\,v-s_3).
\eeq
We conclude via (\ref{fgii}) and (\ref{PBHS}) that ${\cal S}$ is an integral.$\quad\Box$

As in case ($i$) we have $dH\wedge dP_y\wedge dS_1\wedge dS_2=0$, which shows that these four 
observables are not functionally independent. Indeed, we readily find 
\beq\label{reducii}
S_1^2+S_2^2={\cal A}^2+{\cal B}^2=8\,H^3+8a\,H^2\,P_y^2+2c\,H\,P_y^4-2
\eps\la^2\,P_y^6,
\eeq
leading us to consider two different super\-integrable systems, namely
\beq
\label{I1I2}
{\cal I}_1=(H,\,P_y,\,S_1),
\qq\qq 
{\cal I}_2=(H,\,P_y,\,S_2).
\eeq
The Poisson bracket of the two cubic integrals still reduces to a polynomial in the observables $H$ and $P_y$, viz.,
\beq
\{S_1,S_2\}=-8a\,H^2\,P_y-4c\,H\,P_y^3+6\eps\,\la^2\,P_y^5,
\eeq
as in (\ref{reducii}) for $S_1^2+S_2^2$, but this is no longer true for the product   
\beq
S_1\,S_2=\cos(2y)\,{\cal A}\,{\cal B}+\sin(2y)\,\frac{{\cal B}^2-{\cal A}^2}{2}
\eeq
which is a new, independent, observable. This time, the set $(H,P_y,S_1,S_2)$ of first integrals of the geodesic flow does not generate a Poisson algebra.

%%%%%%%%%%%%%%%%%%%%%%%%%%%%%%%%%%%%%%%%%%%%%%%%%%%%%%%%%%%%%%%%%%%%%%%%%%%%%%%%%
\subsection{Transformation of the metric and curvature}
%%%%%%%%%%%%%%%%%%%%%%%%%%%%%%%%%%%%%%%%%%%%%%%%%%%%%%%%%%%%%%%%%%%%%%%%%%%%%%%%%

Returning to the expression (\ref{Dii}) of $D$, let us define the polynomials
\beq
\label{idP}
P=8v D+D'^2,
\qq\qq  
Q=2DD''-D'^2=-P-4(v-s_1)D,
\qq 
Q'=-12\,D,
\eeq
which readily yield the metric 
\beq
\label{metii}
\half G=\rho^2\,\frac{dv^2}{D}+\frac{4D}{P}\,dy^2,
\qq\qq
 -\rho\equiv -\frac{Q}{P}=1+(v-s_1)\,\frac{4D}{P},
\eeq
with the restrictions $D>0$ and $P>0$ that ensure its Riemannian signature. We notice that the scalar curvature is still  given by
\beq
\label{curvii}
R_G=\frac{1}{4Q^3}\Big(2PQ\,W'-(QP'+2PQ')\,W\Big),
\qq 
W\equiv DP'-PD'=8D^2+QD',
\eeq 
showing that Lemma \ref{Lemma1} remains valid. 
% and we will add
\begin{nlem}
\label{Lemma3}
Let $I=(-\nf,v_0)$ be the allowed interval for $v$ where $v_0$ is a simple zero of~$D$. 
%Let it be supposed that 
If for all $v\in\,I$ one has $P(v)>0$ and $Q(v)>0$, then the metric exhibits 
a conical singularity which precludes any manifold structure.  
\end{nlem}
\nin{\bf Proof:} 
Using the relations given in (\ref{idP}), when $v\to\,v_0+$ the metric approximates as
\beq
\half G\approx \frac{4}{D'(v_0)}(dr^2+r^2\,dy^2),
\qq\qq 
r=\sqrt{v-v_0} \to 0+
\eeq
and hence, for this singularity to be apparent, we need to assume $y=\phi\in{\mb S}^1$.

For $v\to -\nf$ we get
\beq
\half G\approx dr^2+r^2\,\left(\frac{d\phi}{3}\right)^2,
\qq\qq 
r=\frac{1}{\sqrt{-v}}\to 0+
\eeq
and we cannot have $\phi/3\in\,{\mb S}^1$ as well. This kind of singularity, called \textit{conical}, rules out a manifold structure. 
$\quad\Box$

For further use we will also prove the general result:
\begin{nlem}
\label{Lemma4}
Assume that the metric
\beq
G=A(v)\,dv^2+B(v)\,d\phi^2,
\qq\qq 
v\in I=[a,b],
\qq
\phi\in{\mb S}^1,
\eeq
be globally defined on a closed manifold $M$. Then its Euler characteristic is given by
\beq\label{gamma}
\chi(M)=\ga(b)-\ga(a),
\qq\qq  
\ga=-\frac{B'}{2\sqrt{A\,B}}.
\eeq
\end{nlem}

\nin{\bf Proof:} Using the orthonormal frame 
\[
e_1=\sqrt{A}\,dv,
\qq\qq 
e_2=\sqrt{B}\,d\phi,
\] 
we find that the connection $1$-form reads $\dst\ \om_{12}=\frac{\ga}{\sqrt{B}}\ e_2$, where $\ga$ is as in (\ref{gamma}). The curvature $2$-form is then given by
\[
R_{12}=d\om_{12}=\frac{\ga'}{\sqrt{AB}}\,e_1\wedge e_2,
\]
from which we get 
\[
\chi(M)=\frac{1}{2\pi}\int_M{}R_{12}=\int_I\,\ga'(v)dv=\ga(b)-\ga(a),
\]
which was to be proved. $\quad\Box$
 
Let us consider now the global properties of these metrics.

%%%%%%%%%%%%%%%%%%%%%%%%%%%%%%%%%%%%%%%%%%%%%%%%%%%%%%%%%%%%%%%%%%%%%%%%%%%%%%%%%
\subsection{The global structure for $\eps=0$}
%%%%%%%%%%%%%%%%%%%%%%%%%%%%%%%%%%%%%%%%%%%%%%%%%%%%%%%%%%%%%%%%%%%%%%%%%%%%%%%%%

In this section we will keep the notation
\[
D(v)=(a-v)(v^2-a^2+c),
\qq\qq 
\De(D)=4c^2(a^2-c),
\]
and organize the discussion according to the values of the discriminant $\De(D)$ of $D$. We will exclude the single case $a=c=0$ since then the scalar curvature vanishes, implying that we loose super\-integrability as explained in the proof of Proposition \ref{Prop2}.

%--------------------------------------------------------------------------------
\subsubsection{First case: $\De(D)=0$}
%--------------------------------------------------------------------------------

We will begin with
\begin{nth} 
There exists no closed manifold for $c=0$ and $a\neq 0$.
\end{nth}
\nin{\bf Proof:} 
We have, in this case,
\beq
D(v)=(a-v)(v^2-a^2),
\quad 
P(v)=(v-a)^4,
\quad 
Q(v)=3(v-a)^3(v-v_*),
\quad 
v_*=-\frac{5}{3}\,a,
\eeq
and the metric writes
\beq
\half G=9\frac{(v-v_*)^2}{(a-v)^4}\frac{dv^2}{-a-v}+\frac{4}{3}\frac{a-v}{(v-v_*)}\,dy^2.\eeq
For $a>0$ we have $D>0$ and $P>0$ iff $v\in I=(-\nf,-a)$; but since $v_*\in I$ we get no manifold structure by Lemma~\ref{Lemma2}. 

For $a<0$ the positivity of $G$ is satisfied for $v\in (-\nf,a)\cap (a,-a)$. In both cases, $a$ is a 
zero of $P$ but we cannot use Lemma \ref{Lemma2} because $Q(a)=0$. In fact, the measure of the sought manifold
\[
\mu_G=12\int \frac{(v-v_*)}{(v-a)^3}\,dv\int\!dy
\]
is divergent (since the integrand blows up at $v=a$), prohibiting a closed manifold. $\quad\Box$

\begin{nth}
There exists no closed manifold for $c=a^2>0$.
\end{nth}
\nin{\bf Proof:} 
We have now
\beq
D(v)=v^2(a-v),
\qq 
P(v)=v^2(v-2a)^2,
\qq 
Q(v)=3v^3(v-v_*),
\quad 
v_*=\frac{4}{3}\,a.
\eeq
For $a<0$ we have $D>0$ and $P>0$ iff either $v\in\,I_1=(2a,a)$ or $v\in I_2=(-\nf,2a)$. 
In the first interval $Q$ has a simple zero $v=v_*$, and $P(v_*)$ and $D(v_*)$ do not vanish; 
in view of Lemma \ref{Lemma1} we get a curvature singularity. As for the second interval, the end-point $v=2a$ is a zero 
of $P$ where $Q(2a)\neq 0$; hence by Lemma \ref{Lemma2}, the sought manifold is not~closed.

For $a>0$ we have $I=(-\nf,a)$. There will be no curvature singularity since $Q$ never vanishes 
for $v\in\,I$. Since $v=a$ is a simple zero of $D$ such that $P(a)$ and $Q(a)$ are non-zero;
we conclude by Lemma \ref{Lemma3}. $\quad\Box$

%--------------------------------------------------------------------------------
\subsubsection{Second case: $\De(D)<0$}
%--------------------------------------------------------------------------------

Here, we have  
\beq
D=(a-v)\Big(v^2+c-a^2\Big),
\qq 
c>a^2,
\qq P=(v-w_-)^2(v-w_+)^2,
\qq 
w_{\pm}=a\pm\sqrt{c},
\eeq
and
\beq
Q=-P+4(a-v)D,
\qq\qq 
Q'=-12\,D.
\eeq

%Let us prove:

\begin{nth}
There exists no closed manifold for $\De(D)<0$.
\end{nth}

\nin{\bf Proof:} 
The positivity of $D$ and $P$ holds for any $\,v\in(-\nf,w_-)\cup \,(w_-,a)$. 
The second interval is excluded since $Q$ is strictly decreasing and the relations
\beq
Q(w_-)=8\,c^{3/2}\,(\sqrt{c}-a)>0,
\qq\qq 
Q(a)=-c^2<0,
\eeq
imply that $Q$ has a simple zero  inside the interval $\,(w_-,a)$, inducing a curvature singularity as 
already explained. This never happens for $\,v\in\,(-\nf,w_-)$ since then $Q(v)>0$. But $w_-$ is a zero of $P$ and $Q(w_-)>0$; we conclude by Lemma \ref{Lemma2}. 
$\quad\Box$ 

%--------------------------------------------------------------------------------
\subsubsection{The case $\De(D)>0$}
%--------------------------------------------------------------------------------

This time, $c<a^2$ and we find
\beq
\barr{l}
\label{DPQv0}
D(v)=(a-v)(v^2-v_0^2),
\qq\qq 
v_0=\sqrt{a^2-c},\\[4mm] 
P(v)=\Big((v-a)^2-c\Big)^2,
\qq\qq 
Q(v)=-P(v)+4(a-v)D(v).
\earr
\eeq
The parameter $c$ can take its values in the set
\[
(-\nf,0)\cup\{ 0\} \cup (0,a^2).
\]
Let us consider first negative values of $c$.

\setcounter{nTh}{0}

\begin{nTh}
\label{Thm1}
If $c\in\,(-\nf,0)$ the superintegrable systems ${\cal I}_1$ and ${\cal I}_2$ given in (\ref{I1I2})  
are globally defined on~${\mb S}^2$.
\end{nTh}

\nin{\bf Proof:} 
First of all, we have $P>0$. The ordering of the zeroes of $D$ 
is $\,-v_0<a<v_0$. This implies two possible intervals ensuring its positivity: either 
$v\in\,(-\nf,-v_0)$ or $v\in\,(a,v_0)$.

The first case is easily ruled out since $Q$ decreases from $Q(-\nf)=+\nf$ to $Q(-v_0)=-P(-v_0)<0$; it thus vanishes in the interval and leads to a curvature singularity.

So let us consider $v\in\,(a,v_0)$. Then $Q(a)=-P(a)=-c^2$ is negative, and since $Q$ is decreasing it will remain strictly negative everywhere on the interval. Putting $v_0=1$ and performing the transformation $G\to2\,G$ for convenience, we end up with the explicit form of the metric, namely
\beq
\label{metepsnul}
G=\rho^2\,\frac{dv^2}{(v-a)(1-v^2)}+4\frac{(v-a)(1-v^2)}{(v^2-2av+1)^2}d\phi^2,
\qq 
v\in (a,1),
\qq
\phi\in\,{\mb S}^1,
\eeq
where
\beq
a\in\,(-1,1),
\qq\qq 
-\rho=1+4\frac{(v-a)^2(1-v^2)}{(v^2-2av+1)^2}.
\eeq
Both end-points are apparent singularities because
%since we have
\beq
G(v\to 1-)\sim\frac{2}{1-a}(dr^2+r^2\,d\phi^2),
\qq\qq 
r=\sqrt{1-v},
\eeq 
and
\beq
G(v\to a+)\sim\frac{4}{1-a^2}(dr^2+r^2\,d\phi^2),
\qq\qq 
r=\sqrt{v-a}.
\eeq
Let us compute the Euler characteristic. Resorting to Lemma \ref{Lemma4}, we find \beq
\ga(v)=\frac{(1-v^2)^2-4(v-a)^2}{Q(v)}
\qq\Longrightarrow\qq 
\chi(M)=\ga(1)-\ga(a)=2,
\eeq
which proves that the manifold is diffeomorphic to ${\mb S}^2$. The measure of this surface is 
%merely 
\beq
\mu_G({\mb S}^2)=\frac{4\pi}{1+a}.
\eeq

Let us investigate now the global status of the integrals $H,P_y,S_1,S_2$. Using (\ref{DPQv0}), and referring to the Riemann uniformization theorem, we can write
\beq
H=\half\left(\Pi^2+P\,\frac{\Pf^2}{4D}\right)=\frac{1}{2\Om^2}\left(P_{\tht}^2+\frac{\Pf^2}{\sin^2\tht}\right)
\eeq
with
\beq
t\equiv\tan\frac{\tht}{2}=\sqrt{\frac{(v-a)P}{(1-v^2)}},
\qq \qq 
\Om=\frac{1-v^2}{P}+v-a,
\eeq
and the conformal factor is indeed $C^{\nf}$ for all $v\in\,[a,1]$.

To ascertain that the previous integrals are globally defined, we will express them in terms of globally defined quantities, e.g., the $\mathrm{SO}(3)$ generators on $T^*{\mb S}^2$, namely
\beq
L_1=-\sin\phi\,P_{\tht}-\frac{\cos\phi}{\tan\tht}\,\Pf,
\qq 
L_2=\cos\phi\,P_{\tht}-\frac{\sin\phi}{\tan\tht}\,\Pf,
\qq L_3=\Pf,
\eeq
and the constrained coordinates
\[
x^1=\sin\tht\,\cos\phi,
\qq\qq 
x^2=\sin\tht\sin\phi,
\qq\qq 
x^3=\cos\tht.
\]

The relation $\Pi=-P_{\tht}/\Om$ and formulas (\ref{Scasii1}) and (\ref{Scasii2}) yield
\beq
S_1=-\frac{L_2}{\Om}\,\left(\Pi^2-Q\,\frac{\Pf^2}{4D}\right)
+x^2\,L_3\,\left(A\,\Pi^2-B\,\frac{\Pf^2}{4D}\right),
\eeq
where the functions $A,B$ of $\tht$ retain the form
\beq
A=\frac{D'-\sqrt{P}\,\cos\tht}{2\sin\tht\,\sqrt{D}},
\qq\qq 
B=\frac{W-Q\sqrt{P}\,\cos\tht}{2\sin\tht\,\sqrt{D}}.
\eeq
The polynomials $P,\,Q$ and $W$ are clearly globally defined, as well as  
the quantities $\Pi^2$ and $\Pf^2/(4D)$ in the Hamiltonian. So, it is sufficient to check that 
the functions $A$ and $B$ are well-behaved near the poles. 

Let us begin with the north-pole 
($v\to a+$ or $\tht\to 0+$) for which we get 
\beq
\left\{
\barr{l}
\dst 
A=\frac{\phi(a)}{2(1-a^2)}-\frac{\sin^2\tht}{4(1-a^2)^2}+O(\sin^4\tht),\\[4mm]
\dst 
B=-\frac{(1-a^2)}{2}\,\phi(a)+\frac{3}{4}\sin^2\tht+O(\sin^4\tht),
\earr
\right. 
\qq 
\phi(a)=a^4-2a^2-2a+1,  
\eeq
while for the south pole ($v\to 1-$ or $\tht \to \pi-$ ) we obtain
\beq
\left\{
\barr{l}
\dst 
A=\frac{\psi(a)}{2(1-a)}-\frac{(1-a)^4}{2}\,\sin^2\tht+O(\sin^4\tht),\\[4mm]
\dst 
B=-2(1-a)\,\psi(a)+6(1-a)^6\,\sin^2\tht+O(\sin^4\tht),
\earr
\right. 
\qq 
\psi(a)=2a^2-4a+1. 
\eeq
We observe that either $\phi(a)$ or $\psi(a)$ may vanish for some $a\in\,(0,1)$, but this 
does not jeopardize the conclusion.

For the other integral, due to the relation 
\beq
S_2=\{\Pf,S_1\}=\frac{L_1}{\Om}\,\left(\Pi^2-Q\,\frac{\Pf^2}{4D}\right)
+x^1\,L_3\,\left(A\,\Pi^2-B\,\frac{\Pf^2}{4D}\right),
\eeq
there is nothing more to check.
$\quad\Box$ 
   
Let us consider now the second case where $c$ vanishes.

\begin{nth} 
For $c=0$ there exists no closed manifold.
\end{nth}

\nin{\bf Proof:} The above functions simplify and read
\beq
D=-(v+a)(v-a)^2,
\qq 
P=(v-a)^4,
\qq 
Q=(3v+5a)(v-a)^3,
\qq 
a\neq 0.
\eeq
For $a>0$ the positivity of $D$ requires $v\in\,I=(-\nf,-a)$, but since $Q$ has a simple zero 
%for 
$\dst v=-\frac{5}{3}\, a\in\,I$, in view of Lemma \ref{Lemma1} there is no manifold structure. 

For $a<0$ either $v\in\,(-\nf,a)$ or $v\in\,(a,-a)$ ensure the positivity of $D$. But in both cases $P$ vanishes for $v=a$, and the measure of the would-be manifold 
\[
\mu_G=12\int \frac{(v-v_*)}{(v-a)^3}\,dv\int\!dy
\]
is divergent, excluding a closed manifold. 
$\quad\Box$
 
The remaining case is $c\in (0,a^2)$. The discussion depends strongly on the sign of $a$. Beginning with $a>0$ we have:
\begin{nth} For $c\in (0,\,a^2)$ and $a<0$ there exists no closed manifold.
\end{nth}

\nin{\bf Proof:}  The two functions $(D,\,P)$ are now
\beq
D(v)=(a-v)(v^2-v_0^2),
\quad 
v_0=\sqrt{a^2-c},
\quad 
P=(v-w_-)^2(v-w_+)^2,
\quad 
w_{\pm}=a\pm\sqrt{c},
\eeq
with the ordering $\ w_-<a<w_+<-v_0\ $. 

The positivity requirements give three possible intervals: 
\[
I_1=(-\nf,w_-),
\qq\qq 
I_2=(w_-,a),
\qq\qq 
I_3=(-v_0,v_0).
\]
\begin{itemize}
\item 
For $v\in I_1$ we notice that $w_-$ is a zero of $P$ for which $Q(w_-)=4(a-w_-)D(w_-)>0$, and 
we conclude by Lemma \ref{Lemma2}.
\item 
For $v\in I_2$ since $Q(w_-)>0$ and $Q(a)=-P(a)<0$, there is a simple zero $v_*$ of $Q$ 
inside $I_2$; hence, by Lemma \ref{Lemma1}, there is no manifold structure.
\item 
For $v\in I_3$ we have $Q(-v_0)=-P(-v_0)<0$ and then $Q$ decreases to $Q(v_0)$; it thus never 
vanishes and $P>0$ in $I_3$, opening the possibility of a manifold structure.
\end{itemize}
Putting $v_0=1$ and computing the metric brings us back to (\ref{metepsnul}). 
$\quad\Box$

For $a>0$ we have:
\begin{nth} 
For $c\in (0,\,a^2)$ and $a>0$ there exists no closed manifold.
\end{nth}

\nin{\bf Proof:} The zeros of $D$ and $P$ interlace as follows  
$\ w_-<-|a|<-v_0<w_+<0<v_0\ $ giving four possible intervals
\[
I_1=(-\nf,w_-),
\qq 
I_2=(w_-,-|a|),
\qq 
I_3=(-v_0,w_+),
\qq I_4=(w_+,v_0).
\]

\begin{itemize}
\item 
For $v\in\,I_1=(-\nf,w_-)$, and since $w_-$ is a zero of $P$, we use Lemma \ref{Lemma2}.
\item 
If $v\in\,I_2=(w_-,-|a|)$, then $Q$ is strictly decreasing with
\[
Q(w_-)=4(-w_-+a)\,D(w_-)>0
\qq\mbox{and}\qq 
Q(-|a|)=-P(-|a|)<0,
\]
so that $Q$ has a simple zero in $I_2$; thanks to Lemma \ref{Lemma1}, there is no manifold structure.
\item 
For $v\in\,I_3=(-v_0,w_+)$ or $v\in\,I_4=(w_+,v_0)$, since $w_+$ is a zero of $P$ we invoke again Lemma \ref{Lemma2}. 
$\quad\Box$
\end{itemize}

%%%%%%%%%%%%%%%%%%%%%%%%%%%%%%%%%%%%%%%%%%%%%%%%%%%%%%%%%%%%%%%%%%%%%%%%%%%%%%%%%
\subsection{The global structure for $\eps\neq 0$}
%%%%%%%%%%%%%%%%%%%%%%%%%%%%%%%%%%%%%%%%%%%%%%%%%%%%%%%%%%%%%%%%%%%%%%%%%%%%%%%%%

Let us begin with
\begin{nth}
\label{Prop12}
If $\De(D)=0$ the super\-integrable system is never globally defined on a closed manifold.
\end{nth}

\nin{\bf Proof:} We may have either $D(v)=(v_0-v)^3$ or $D(v)=(v_0-v)(v-v_1)^2$ with $v_0\neq v_1$. 

The first case is ruled out as in Proposition \ref{Prop2} since the metric is of constant curvature. 

In the second case we have
\beq
\left\{
\barr{l}
P(v)=(v-v_1)^2\,p(v),
\qq\qq
p(v)=v^2-2(2v_0+3v_1)v+(2v_0+v_1)^2,\\[4mm]
\dst
Q(v)=3(v-v_1)^3(v-v_*),
\qq 
v_*=v_0+\frac{v_0-v_1}{3}
\earr
\right.
\eeq

Let us first consider the case $v_0<v_1$. Then $D$ is positive iff $v\in I=(-\nf,v_0)$. If $\De(p)<0$ then $P>0$ for all $v\in I$. But, since $v_*<v_0$, 
there will be a curvature singularity inside $I$. If $\De(p)$ vanishes, we get 
$p(v)=(v-w_0)^2$ and either $w_0=v_0<0$ or $w_0=2v_0<0$. In the first case there will be a curvature 
singularity at $\dst v_*=\frac{4}{3}\,v_0\in I$ while, in the second case, the positivity interval becomes 
$(-\nf,w_0)$; since $v=w_0$ is a zero of $P$ we use Lemma~\ref{Lemma2}. If $\De(p)<0$ we have 
two real zeroes and $p(v)=(v-w_-)(v-w_+)$. The interval of positivity becomes 
$I=(-\nf,v_0)\cap (w_-,w_+)$ and since at least one of its end-points will correspond to a zero of 
$P$ we conclude by Lemma 2.

Let us then consider the other case $v_0>v_1$. Then $D$ is positive iff $v\in I=(-\nf,v_0)$. If  $\De(p)<0$ then $P>0$ for all $v\in I$, and we conclude by Lemma \ref{Lemma3}. If $\De(p)=0$ we get 
$p(v)=(v-w_0)^2$, and either $w_0=v_0>0$ or $w_0=2v_0>0$. In the first case we remain with 
$v\in(-\nf,v_0)$ and end up with a conical singularity for $v\to -\nf$; in the second case $v\in(w_0,v_0)$ where $w_0$ is a zero of $P$, which excludes closedness by Lemma \ref{Lemma2}. If $\De(p)>0$ we have two real zeroes and $p(v)=(v-w_-)(v-w_+)$. The interval of positivity becomes $I=(-\nf,v_0)\cap (w_-,w_+)$ and 
at least one of its end-points will correspond to a zero of $P$; we conclude by Lemma \ref{Lemma2}. $\quad\Box$ 

Let us proceed to

\begin{nth} 
\label{Prop13}
If $\De(D)<0$ the super\-integrable systems are never globally defined on a 
closed manifold.
\end{nth}

\nin{\bf Proof:} In this case, we can write
\beq
D(v)=(v_0-v)[(v-a)^2+b^2],
\quad 
b\neq 0,
\qq Q(v)=-P(v)+4(v_0+2a-v)D(v),
\eeq
and
\beq
\label{eqpii}
P(v)=p(v)^2-16a[(v_0+a)^2+b^2]v,
\qq 
p(v)=v^2-2(v_0+2a)v-a^2-b^2-2av_0.
\eeq
We have $D>0$ iff $v\in I=(-\nf,v_0)$. Let us also notice that $\De(P)$ and $\De(Q)$ being negative, $P$ and $Q$ will have two simple real zeroes. Since $Q(v_0)<0$, then $Q$ will have a simple zero $v_*<v_0$. 

If $a=0$ we have $p(v)=(v-w_-)(v-w_+)$, with the ordering $w_-<w_+$; hence $P$ is always positive, but its zeroes may change the interval for $v$: if $w_-<v_*$ the interval for~$v$ becomes $(w_-,\,v_0)$ and then $v_*$ is a curvature singularity inside this interval; if $w_->v_*$ the interval 
for $v$ becomes $(w_-,\,v_0)$ for which Lemma \ref{Lemma2} applies.

If $a>0$, the relation (\ref{eqpii}) tells us that both roots of $P$ must be positive and, since $P(v_0)=\big((v_0-a)^2+b^2\big)^2>0$, they must lie to the right of $v_0$. The interval for $v$ remains $(-\nf,v_0)$ and we conclude by Lemma \ref{Lemma3}.

If $a<0$ both roots of $P$ ordered as $w_-<w_+$ must be negative and to the left of $v_0$. The 
positivity of $P$ will reduce the interval of $v$ either to $(-\nf,w_-)$ or to $(w_+,v_0)$ and in 
both cases Lemma \ref{Lemma2} allows us to conclude. $\quad\Box$

Let us end up this section with:

\begin{nTh}\label{Thm2}
If $\De(D)>0$ one can put $D(v)=-(v-v_0)(v-v_1)(v-v_2)$ with $v_0 < v_1 < v_2$; the super\-integrable systems $\ {\cal I}_1$ and $\ {\cal I}_2$ given by (\ref{I1I2}) are indeed globally defined on ${\mb S}^2$ iff~$v_0+v_2>0$. 
\end{nTh}

\nin{\bf Proof:} Let us define the symmetric polynomials of the roots $s_1,s_2,s_3$ by
\[D(v)=-(v-v_0)(v-v_1)(v-v_2)=-v^3+s_1\,v^2-s_2\,v+s_3.\]
The function $D$ is positive iff either $v\in(-\nf,v_0)$ or $v\in(v_1,v_2)$. 
Let us first study the polynomial $Q=3v^4-4s_1\,v^3+\cdots$. Since  
\[
\De(Q)=-6912(v_1-v_0)^4(v_2-v_0)^4(v_2-v_1)^4<0
\] 
we conclude that $Q$ has two simple real zeroes. For $v\in\,(v_0,v_1)$ the relation $Q'=-12\,D$  shows that 
$Q$ increases from $Q(v_0)=-(v_0-v_1)^2(v_0-v_2)^2$ to $Q(v_1)=-(v_1-v_0)^2(v_1-v_2)^2$; it then
decreases to $Q(v_2)=-(v_2-v_0)^2(v_2-v_1)^2$ so that $Q$ is strictly negative for all $v\in\,(v_0,v_2)$ and, since $Q(\pm\nf)=+\nf$, it will have a simple zero at $v=v_*<v_0$ and 
%the other one 
at $v=\wti{v}_*>v_2$, with the relation $\dst v_*+\wti{v}_*=\frac{4}{3}\,s_1.$

Let us come back to the first positivity interval for $D$ which is $I=(-\nf,v_0)$. As we have already seen, $Q$ has a simple zero $v_*\in I$. Let us prove that $P(v_*)>0$ which will be sufficient to 
ascertain, thanks to Lemma \ref{Lemma1}, that $v=v_*$ is a curvature singularity. To this end we use the relation
\beq
P(v)=-Q(v) +4(s_1-v)D(v)
\qq\Longrightarrow\qq 
P(v_*)=4(s_1-v_*)\,D(v_*).
\eeq
Since $v_*<v_0$ we have $D(v_*)>0$ and 
\[
s_1-v_*=\wti{v}_*-\frac{s_1}{3}>v_2-\frac{s_1}{3}=\frac{2v_2-v_0-v_1}{3}>0.
\]

Let us now consider the second positivity interval for $D$ which is $I=(v_1,v_2)$. We find it convenient to define new parameters by
\beq
d=\frac{v_2-v_1}{2}>0,
\qq\qq 
l=\frac{v_1+v_2+2v_0}{v_2-v_1}\in{\mb R},
\qq\qq 
m=\frac{v_1+v_2-2v_0}{v_2-v_1}>1,
\eeq
and a new coordinate, $x$, by
\beq
v=d\Big(x+\frac{l+m}{2}\Big),
\qq\qq 
x\in \ I=\,[-1,+1].
\eeq
Since $d>0$ we will set $d=1$. It follows that
\beq
\left\{
\barr{ll}
D=(x+m)(1-x^2), 
&  \\[4mm]
P=\Big(L_+\,(1-x^2)+2(m+x)\Big)\Big(L_-\,(1-x^2)+2(m+x)\Big), 
& 
\qq L_{\pm}=l\pm\sqrt{l^2-1},\\[4mm] 
Q=3x^4+4mx^3-6x^2-12mx-4m^2-1,
& 
\qq 
Q'=-12\,D,
\earr
\right.
\eeq
and the metric (again up to the change $G\to 2\,G$) reads now
\beq
G=\rho^2\,\frac{dx^2}{D}+\frac{4D}{P}\,d\phi^2,
\qq\qq 
\rho=\frac{Q}{P}.
\eeq

For $x\in\,I$ the polynomial $Q$ decreases from $Q(-1)=-4(m-1)^2$ to $Q(1)=-4(m+1)^2$ forbidding any curvature singularity. It remains to check the positivity of $P$. Its factorized expression shows that for $l\in\,[-1,1)$ it has no real root. For $l\geq 1$ it has four simple real roots 
which lie outside $I$, and for $l<-1$ two of its real roots are still outside $I$, the remaining two $x_-<x_+$ being contained in $I$. It follows that $I$ may be reduced to any of the 
intervals
\[
I_1=(-1,x_-)
\qq 
\mbox{or}
\qq 
I_2=(x_-,\,x_+)
\qq 
\mbox{or}\qq 
I_3=(x_+,1).\]
Now, at least one end-point is a zero of $P$, and by Lemma \ref{Lemma3}, the expected manifold is not closed. 
So far, we have proved that a manifold can exists iff $\ l\in\,(-1,+\nf)$, which translates as $v_0+v_2>0$.
 
Let us study the the behavior of the metric at the end-points of $I$ by setting $x=\cos\vartheta$ with $\vartheta\in(0,\pi)$. We find that
\beq
G(\vartheta\to 0+)\approx \frac{1}{m+1}(d\vartheta^2+\sin^2\vartheta\,d\phi^2),
\qq 
G(\vartheta\to \pi-)\approx \frac{1}{m-1}(d\vartheta^2+\sin^2\vartheta\,d\phi^2),
\eeq
and $\vartheta=0,\pi$ are indeed apparent singularities. From Lemma \ref{Lemma4} we get
\beq
\ga=-\frac{W}{Q\,\sqrt{P}}\qq W=-(x^2+2x-1+2m)(x^2-2x-1-2m)(x^2+2mx+1)
\eeq
which gives
\[
\chi(M)=\ga(1)-\ga(-1)=2,
\]
so that the manifold is actually $M\cong{\mb S}^2$.

Returning to the integrals, we will define once more
\beq
H=\half\left(\Pi^2+P\,\frac{\Pf^2}{4D}\right)=\frac{1}{2\Om^2}\left(P_{\tht}^2+\frac{\Pf^2}{\sin^2\tht}\right),
\eeq
which leads to the relations  
\beq
\Om^2\,\sin^2\tht=\frac{4D}{P},
\qq
\frac{d\tht}{\sin\tht}=\frac{F(x)}{(1-x^2)}\,dx,
\qq  
F(x)=\frac{Q(x)}{2(m+x)\sqrt{P(x)}},
\eeq
from which we deduce
\beq
t\equiv\tan\frac{\tht}{2}=\exp{\left(\int_0^x\,\frac{F(u)}{(1-u^2)}\,du\right)}.
\eeq
We need first to check the behavior of the conformal $\Omega$ factor at the north pole for $x\to 1-$. We have
\beq
\label{polenord}
t=\sqrt{1-x}\ T_N(x),
\quad
T_N(x)=\exp{(U(x))},
\quad
U(x)=\int_0^x\left(\frac{F(u)}{1+u}-\frac{F(1)}{2}\right)\frac{du}{1-u},
\eeq
so that $T_N$ is $C^{\nf}$ in a neighborhood of $x=+1$. This implies that
\beq
\Om^2=\frac{(1+t^2)^2(m+x)(1+x)}{P(x)\,T_N^2(x)}
\eeq
is also $C^{\nf}$ in a neighborhood of $x=+1$. At the south pole, i.e., for $x\to\,-1+$ a similar 
argument works.

The expression of $S_1$, in view of $\Pi=P_{\tht}/\Om$, is now the following: 
\beq
S_1=\frac{L_2}{\Om}\,\left(-\Pi^2+Q\,\frac{\Pf^2}{4D}\right)
+x^2\,L_3\,\left(A\,\Pi^2-B\,\frac{\Pf^2}{4D}\right)
\eeq
with
\beq
A=\frac{D'+\sqrt{P}\,\cos\tht}{2\sin\tht\,\sqrt{D}},
\qq\qq 
B=\frac{W+Q\sqrt{P}\,\cos\tht}{2\sin\tht\,\sqrt{D}},
\eeq
giving at the north pole:
\beq
\left\{
\barr{l}
\dst 
A=\frac{1}{\sqrt{2(m+1)}}\frac{(l+m+2)}{2\,T_N(1)}+O(\sin^2\tht),\\[6mm] 
\dst 
B=-\Big(2(m+1)\Big)^{3/2}\frac{(l+m+2)}{2\,T_N(1)}+O(\sin^2\tht),
\earr\right.\eeq
where the leading coefficients never vanish since $l+m>0$. 

%\goodbreak

To analyze the behavior of $S_1$ at south pole let us define
\beq
\label{polesud}
t=\frac{1}{\sqrt{1+x}\, T_S(x)},
\;
%\quad 
T_S(x)=\exp{(-V(x))},
\;
%\quad
V(x)=\int_0^x\left(\frac{F(u)}{1-u}-\frac{F(-1)}{2}\right)\frac{du}{1+u},
\eeq
from which we deduce
\beq
\left\{
\barr{l}
\dst 
A=-\frac{1}{\sqrt{2(m-1)}}\frac{(l+m-2)}{2\,T_S(-1)}+O(\sin^2\tht),\\[6mm] 
\dst 
B=(2(m-1))^{3/2}\frac{(l+m-2)}{2\,T_S(-1)}+O(\sin^2\tht),
\earr\right.\eeq
which are well-behaved. For $l+m=2$ the power series expansions begin with 
$\sin^2\tht$, a  
possibility already observed in the proof of Theorem \ref{Thm1}. 

As to the integral $S_2$, the argument given in the proof of Theorem \ref{Thm1}  works here just as well. 
$\quad\Box$

%%%%%%%%%%%%%%%%%%%%%%%%%%%%%%%%%%%%%%%%%%%%%%%%%%%%%%%%%%%%%%%%%%%%%%%%%%%%%%%%%
\subsection{{Comparison with the results of Matveev and Shevchishin}}
\label{newSubsection}
%%%%%%%%%%%%%%%%%%%%%%%%%%%%%%%%%%%%%%%%%%%%%%%%%%%%%%%%%%%%%%%%%%%%%%%%%%%%%%%%%

{In \cite{ms} it was stated in Theorem 6.1 that the metric
\[g=\frac{dx^2+dy^2}{h_x^2},\qq\qq h_x=\frac{dh}{dx},\]
where $h$ is a solution of the differential equation (3,ii) with
\beq
\mu=1,
\qq A_0=1,
\qq A_1=0,
\qq A_3=A_4=A_e>0\qq\mbox{and}\qq A_2\in{\mb R},
\eeq
is globally defined on $S^2$. As we will show in what follows, our results are partly in agreement 
with this Theorem 6.1.}

{Let us first write again the metric in our $(v,\phi)$ coordinates:
\beq\label{vds}
g=\frac{Q^2}{P^2}\,\frac{dv^2}{D}+\frac{4D}{P}\,d\phi^2,\qq\quad \phi\in{\mb S}^1,
\eeq
where $Q$ and $P$ are deduced from the knowledge of $D$ by the relations given in (\ref{idP}). So 
to be able to compare these metrics we have first to notice that 
\[h=\frac{D'}{2\sqrt{2D}}\qq\quad D'=\frac{dD}{dv}, \qq\qq h_x=u=\sqrt{h^2+v}=\sqrt{\frac{P}{8D}}>0,
\]
and that
\[y=\phi,\qq\qq \frac{dx}{dv}=\frac{dx}{dh}\frac{dh}{dv}=\frac 1{h_x}\frac{dh}{dv}=\frac{Q}{2D\sqrt{P}}.\]}
{Let us notice that to be riemannian  the metric (\ref{vds}) requires  $D>0$ and $P>0$ and for the 
transformation $v \to x$ to be locally bijective we need $Q$ to have a fixed sign.}

{Under the hypotheses of Theorem 6.1 we have, in our notation,
\[\eps=1,\qq\qq a=-A_2,\qq\qq\la=2A_e,\]
which gives
\begin{equation}
D=-(v+A_2)(v^2-A_2^2+c)-8A_e^2,
\label{newD}
\end{equation}
where $c$ is a constant of integration which does not appear in the proof of Theorem 6.1 and which can be freely chosen.}

{The discriminant of $D$ is 
\[\De(D)=-27\,\xi^2+4A_2(8A_2^2-9c)\xi+4c^2(A_2^2-c),
\qq\qq \xi=8A_e^2,
\]
and the crucial point is that the sign of this discriminant is undefined. If $\De(D)>0$, then Theorem 6.1 of Matveev and Shevchishin agrees with our Theorem 2, and the metric is 
indeed globally defined on ${\mb S}^2$. Nevertheless, if $\De(D)\leq 0$ our Propositions \ref{Prop12} and \ref{Prop13} show that either curvature singularities or conical singularities rule out any closed manifold. Let us mention that we have also found a metric globally defined on ${\mb S}^2$ for $\epsilon=0$ (see Theorem \ref{Thm1}), a case which has not been studied in \cite{ms}.}

%%%%%%%%%%%%%%%%%%%%%%%%%%%%%%%%%%%%%%%%%%%%%%%%%%%%%%%%%%%%%%%%%%%%%%%%%%%%%%%%%
%%%%%%%%%%%%%%%%%%%%%%%%%%%%%%%%%%%%%%%%%%%%%%%%%%%%%%%%%%%%%%%%%%%%%%%%%%%%%%%%%
\section{The affine case}\label{affineSection}
%%%%%%%%%%%%%%%%%%%%%%%%%%%%%%%%%%%%%%%%%%%%%%%%%%%%%%%%%%%%%%%%%%%%%%%%%%%%%%%%%
%%%%%%%%%%%%%%%%%%%%%%%%%%%%%%%%%%%%%%%%%%%%%%%%%%%%%%%%%%%%%%%%%%%%%%%%%%%%%%%%%

%\vspace{5mm}

In this last case, we will prove that there is no closed manifold for the metric. However, since the analysis is much simpler we will determine the metrics globally defined either on ${\mb R}^2$ or on ${\mb H}^2$.

%%%%%%%%%%%%%%%%%%%%%%%%%%%%%%%%%%%%%%%%%%%%%%%%%%%%%%%%%%%%%%%%%%%%%%%%%%%%%%%%%
\subsection{The metric}
%%%%%%%%%%%%%%%%%%%%%%%%%%%%%%%%%%%%%%%%%%%%%%%%%%%%%%%%%%%%%%%%%%%%%%%%%%%%%%%%%

The differential equation and the metric are
\beq
\label{odecase3}
h_x\Big(h_x^2+A_1\,h+A_2\Big)=A_3\,x+A_4,
\qq\qq 
G=\frac{dx^2+dy^2}{h_x^2},
\eeq
see (\ref{odeglob}, $iii$) and (\ref{Ghhx}).
Differentiating the equation for $h$ gives
\[
\Big(3\,h_x^2+A_1 h+A_2\Big)h_{xx}+A_1\,h_x^2=A_3,
\]
and regarding again $u=h_x$ as a function of the new variable $h$, we rewrite the previous equations as
\[
u(3u^2+A_1 h+A_2)\frac{du}{dh}=A_3-A_1 u^2.
\]
Considering the inverse function $h(u)$ we end up with a linear ode, namely
\beq
(A_3-A_1 u^2)\frac{dh}{du}-A_1u\,h=u(3u^2+A_2).
\eeq
Two cases have to be considered:

\brm
\item 
If $A_1=0$ then $A_3$ cannot vanish; positing 
%we get 
$\dst \mu=\frac{3u^2+A_2}{A_3}$,
the original 
variable, $x$, and the metric, $G$, are now given by 
\beq
\label{met3}
dx=\mu\,du,
\qq 
\mu=\frac{1}{u}\,\frac{dh}{du}
\qq\Longrightarrow\qq 
G=\frac{1}{u^2}\left(\mu^2\,du^2+dy^2\right).
\eeq
Interestingly, the relations
\[
h=h_0+\frac{A_2}{2A_3}u^2+\frac{3}{4}\frac{u^4}{A_3},
\qq\qq 
A_3\,x+A_4=A_2\,u+u^3
 \]
show that we have integrated the ode (\ref{odecase3}) by expressing the function 
$h$ and the variable~$x$ parametrically in terms of $u$.
\item 
If $A_1\neq 0$ we can set $A_1=1$ and, by a shift of $h$, we may put $A_2=0$. To simplify matters, we will perform the following rescalings: $y\to2y$, and $G\to\frac{1}{4}G$. This time, we will define 
\[
-2\mu=\frac{1}{u}\,\frac{dh}{du}
\qq\Longrightarrow\qq 
G=\frac{1}{u^2}\left(\mu^2\,du^2+dy^2\right),
\]
and we get two possible solutions for $\mu$:
\beq\label{mu1mu2}
\mu=1+\frac{C}{(u^2-A_3)^{3/2}}
\qq\mbox{or}\qq 
\mu=1+\frac{C}{(A_3-u^2)^{3/2}},
\eeq
where $C$ is a real constant of integration.
\erm

%%%%%%%%%%%%%%%%%%%%%%%%%%%%%%%%%%%%%%%%%%%%%%%%%%%%%%%%%%%%%%%%%%%%%%%%%%%%%%%%%
\subsection{Global structure for vanishing $\mathbf{A_1}$}
%%%%%%%%%%%%%%%%%%%%%%%%%%%%%%%%%%%%%%%%%%%%%%%%%%%%%%%%%%%%%%%%%%%%%%%%%%%%%%%%%

We have just seen that $\dst \mu=\frac{3u^2+A_2}{A_3}$, and must thus discuss  two cases separately:
\brm
\item First case: $A_2=0$, then we can pose $\mu=2u^2$.
\item Second case: $A_2\neq 0$, then we can pose $\mu=1+au^2$.
\erm

%--------------------------------------------------------------------------------
\subsubsection{The case $\mathbf{A_2=0}$}
%--------------------------------------------------------------------------------

The relation (\ref{met3}) and the change $u\to v=u^2$ yield the 
metric and Hamiltonian, viz.,
\[
G=dv^2+\frac{dy^2}{v} 
\qq\Longrightarrow\qq 
H=\half(P_v^2+v\,P_y^2),
\]
while the cubic integrals read now 
\beq
S_1=\frac{2}{3}\,P_v^3+P_y^2(v\,P_v+\frac{y}{2}\,P_y)
\eeq
and
\beq
\label{reducS2}
S_2=y\,S_1-\left(\frac{y^2}{4}+\frac{v^3}{9}\right)P_y^3-\frac{2}{3}\,v^2\,H\,P_y.
\eeq
This last relation shows that $S_2$ is not algebraically independent, and that the  
super\-integrable system we are considering is just generated by $(H,\,P_y,\,S_1)$. 
Let us mention, for completeness, the following Poisson brackets, namely
\beq
\{P_y,S_1\}=\half\,P_y^3,
\qq \{P_y,S_2\}=S_1,
\qq 
\{S_1,S_2\}=\frac{3}{2}\,S_2\,P_y^2.
\eeq

%Let us begin with:

\begin{nth} For $A_2=0$ the super\-integrable system $(H,P_y,S_1)$ is not 
globally defined.
\end{nth}
\nin{\bf Proof:} The Riemannian character of the metric requires $v>0$ and $y\in{\mb R}$. If 
this metric were defined on a manifold, the scalar curvature would be everywhere defined. An easy computation gives for result $\dst R_G=-\frac{3}{2v^2}$ which is singular for $v\to 0+$.$\quad\Box$

%\nin Let us now consider:

%--------------------------------------------------------------------------------
\subsubsection{The case $\mathbf{A_2\neq 0}$}
%--------------------------------------------------------------------------------

We have now the Hamiltonian
\beq
2H=u^2\left(\frac{P_u^2}{\mu^2}+P_y^2\right),
\qq 
u>0,
\quad 
y\in{\mb R},
\qq 
\mu=1+au^2,
\quad 
a\in{\mb R},
\eeq
and the cubic integrals are respectively
%on the one hand
\beq
S_1=\frac{2a}{3}\left(\frac{u}{\mu}\,P_u\right)^3+P_y\Big(u\,P_u\,P_y+y\,P_y^2\Big)
\eeq
and 
%on the other hand
\beq\label{S2cas2}
S_2=y\,S_1-\half\Big(y^2+u^2(1+au^2/3)^2\Big)P_y^3-\frac{a}{3}u^2(2+au^2)HP_y.
\eeq
The non-trivial Poisson brackets of the observables are then given by
\beq
\{P_y,S_1\}=P_y^3,
\qq 
\{P_y,S_2\}=S_1,
\qq 
\{S_1,S_2\}=3\,S_2\,P_y^2\,+4\,P_y^3\,H
+\frac{16}{3}\,a\,P_y\,H^2.\eeq

%Let us prove:

\begin{nth}\label{cas12}
For $A_2\neq 0$ the super\-integrable system $(H,P_y,S_1)$
\brm
\item is not globally defined for $a<0$,
\item is trivial for $a=0$,
\item is globally defined on $M\cong{\mb H}^2$ for $a>0$.  
\erm
\end{nth}

\nin{\bf Proof:} The scalar curvature reads now
\[
R_G=-\frac{2}{\mu^3}\,(1+3au^2),
\qq\qq%\qq 
u>0,
\quad 
y\in{\mb R}.
\]

If $a<0$ it is singular for $\dst u_0=|a|^{-1/2}$, and the system cannot be defined on a 
manifold. 

For $a=0$ the metric reduces to the canonical metric
\[
G({\mb H}^2,{\rm can})=\frac{du^2+dy^2}{u^2}.\]
of the hyperbolic plane ${\mb H}^2$. As a consequence of Thompson's theorem, which has been recalled above, $S_1$ and $S_2$ are reducible. Of course the set $(H,\,P_y)$ still remains 
an integrable system but it is trivial in the sense that it is no longer super\-integrable.

Let us examine the last case for which $a>0$. The change of coordinates 
\[
t=u\left(1+\frac{a}{3}\,u^2\right)
\;\longmapsto\;
%\quad\mapsto\quad
%\qq \Rightarrow \qq   
u=\frac{\xi^{1/3}}{a}-\xi^{-1/3},
\qq 
\xi(t)=\frac{3}{2}\,a^2\,t+\sqrt{a^3+\frac{9}{4}\,a^4\,t^2},
\] 
implies that $u(t)$ is $C^{\nf}$ for all $t\geq 0$.

In these new coordinates the metric becomes
\beq
G=\Om^2\,\frac{dt^2+dy^2}{t^2}=\Om^2\,G({\mb H}^2,{\rm can}),
\qq%\qq 
\Om(t)=1+\frac{a}{3}\,u^2(t),
\qq 
t>0,
\quad 
y\in{\mb R},
\eeq
and, since $\Om$ never vanishes, it is globally conformally related to the canonical metric of the hyperbolic plane, $M\cong{\mb H}^2$. 

Using the generators of $\mathrm{sl}(2,{\mb R})$ on $T^*{\mb H}^2$ (given in the Appendix) allows us to write the Hamiltonian in the new guise
\beq
H=\frac{t^2}{2\,\Om^2}\Big(P_t^2+P_y^2\Big)=\frac{1}{2\,\Om^2}\Big(M_1^2+M_2^2-M_3^2\Big).\eeq
The relations
\[P_y=M_2+M_3\qq\mbox{and}\qq t\,P_t=\frac{M_1-x^1\,P_1}{1+(x^1)^2}\]
show that 
\beq 
S_1=\frac{2a}{3}\,\left(\frac{t\,P_t}{\Om}\right)^3
+P_y^2\left(\mu\,\frac{t\,P_t}{\Om}+y\,P_y\right),
\qq%\qq 
\mu(t)=1+a\,u^2(t),
\qq 
a>0,
\eeq
is globally defined on $M$. The same is true for $S_2$ (see the relation (\ref{S2cas2})).
$\quad\Box$

%%%%%%%%%%%%%%%%%%%%%%%%%%%%%%%%%%%%%%%%%%%%%%%%%%%%%%%%%%%%%%%%%%%%%%%%%%%%%%%%%
\subsection{Global structure for non-vanishing $\mathbf{A_1}$}
%%%%%%%%%%%%%%%%%%%%%%%%%%%%%%%%%%%%%%%%%%%%%%%%%%%%%%%%%%%%%%%%%%%%%%%%%%%%%%%%%

In the formula (\ref{mu1mu2}) let us change $A_3\to a$. We have, again, two cases to consider according to $\eps={\rm sign}(u^2- a)$.

%--------------------------------------------------------------------------------
\subsubsection{First case: $\mathbf{\eps=+1}$}
%--------------------------------------------------------------------------------

The metric and the Hamiltonian are given by
\beq
G=\frac{1}{u^2}\Big(\mu^2\,du^2+dy^2\Big),
\qq%\qq 
H=\frac{u^2}2\left(\frac{P_u^2}{\mu^2}+P_y^2\right),
\qq%\qq 
u^2-a>0,
\quad 
y\in \mb{R},
\eeq
where
\[
\mu=1+\frac{C}{(u^2-a)^{3/2}}.
\]
The cubic integrals are then
\beq\label{S1cas2}
S_1=\left(\frac{u}{\mu}\,P_u\right)^3+u(u^2-a)\,P_u\,P_y^2-ay\,P_y^3+2y\,H\,P_y
\eeq
and
\beq\label{S2cas3}
S_2=y\,S_1+\half\left(a(u^2+y^2)-\frac{2Cu^2}{\sqrt{u^2-a}}+\frac{C^2}{u^2-a}\right)P_y^3
-\left(u^2+y^2-\frac{2C}{\sqrt{u^2-a}}\right)H\,P_y.
\eeq

The case $C=0$ corresponds to the canonical metric on ${\mb H}^2$, and, as already explained in 
Proposition \ref{cas12}, the system becomes trivial. 

In the following developments, we will discuss the global properties of our super\-integrable system 
according to the sign of~$C \neq 0$, rescaling it to $\pm 1$. 
%Let us prove:

\begin{nth}\label{Cneg} For $C=-1$ the super\-integrable system  
$(H,P_y,S_1)$ is globally defined iff $a<0$ and $|a|>1$, in which case the manifold is $M\cong{\mb H}^2$.
\end{nth}
\nin{\bf Proof:} The scalar curvature is
\beq
R_G=-\frac{2}{\mu^3}\left(1+\frac{(2u^2+a)}{(u^2-a)^{5/2}}\right).
\eeq 
For $a\geq 0$ we must have $u>\sqrt{a}$ and $R_G$ will be singular for $u_0=\sqrt{a+1}$. 
For $a<0$ we must have $u>0$. Then the curvature is singular for $u_0=\sqrt{1-\rho}$ if 
$\,\rho=|a|\leq 1$. 
However for $\rho>1$ the function $\mu$ no longer vanishes and the curvature remains continuous for all $u\geq 0$. The metric then reads 
\beq
G=\frac{1}{u^2}\Big(\mu^2\,du^2+dy^2\Big),
\qq\qq 
\mu=1-\frac{1}{(\rho+u^2)^{3/2}},
\qq 
u>0,
\quad 
y\in{\mb R}.
\eeq
Let us define the new variable
\[
t=u\left(1-\frac{1}{\rho\sqrt{\rho+u^2}}\right),
\qq\qq 
u \in [0,+\nf)
\longmapsto 
t \in [0,+\nf).
\]

Since $\dst \mu=\frac{dt}{du}$ never vanishes, the inverse function $u(t)$ 
is $C^{\nf}([0,+\nf))$ and the metric can be written as
\beq
G=\Om^2\,G(\mb{H}^2,{\rm can}),
\qq\qq 
\Om(t)=1-\frac{1}{\rho\sqrt{\rho+u^2(t)}},
\qq\rho>1,
\eeq
where the conformal factor $\Om(t)$ is $C^{\nf}$ and never vanishes: the manifold is again $M\cong{\mb H}^2$. 

The first cubic integral 
\beq
S_1=\left(\frac{t\,P_t}{\Om}\right)^3+\mu(t)(\rho+u^2(t))\left(\frac{t\,P_t}{\Om}\right)P_y^2
+\rho\,y\,P_y^3+2y\,H\,P_y
\eeq
is therefore globally defined (with same argument as in the proof of Proposition (\ref{cas12})),
and~(\ref{S2cas3}) gives
\[
S_2=y\,S_1+\half\left(-\rho(u^2+y^2)+\frac{2u^2}{\sqrt{\rho+u^2}}+\frac{1}{\rho+u^2}\right)P_y^3
-\left(u^2+y^2+\frac{2}{\sqrt{\rho+u^2}}\right)H\,P_y,
\]
showing that this is also true for $S_2$.
$\quad\Box$

%Let us prove:

\begin{nth} For $C=+1$ the super\-integrable system $(H,P_y,S_1)$ is 
globally defined either if $a>0$ and the manifold is $M\cong{\mb R}^2$, or if $a<0$ and 
%the manifold is 
$M\cong{\mb H}^2$. 
\end{nth}

\nin{\bf Proof:} 
The metric reads now
\beq
G=\frac{1}{u^2}\Big(\mu^2\,du^2+dy^2\Big),
\qq\qq 
\mu=1+\frac{1}{(u^2-a)^{3/2}}.
\eeq
Consider first the case $a>0$ for which $u>\sqrt{a}$. Let us define the new coordinate
\[
t=u\left(1-\frac{1}{a\sqrt{u^2-a}}\right),
\qq\qq 
u \in (\sqrt{a},+\nf)\longmapsto t \in {\mb R}.
\]

Since, again, $\dst \mu=\frac{dt}{du}$ does not vanish $u(t)$ is $C^{\nf}({\mb R})$, 
and the metric
\beq
G=\frac{dt^2+dy^2}{u^2(t)},
\qq\qq 
t\in{\mb R},
\quad 
y\in{\mb R},
\eeq
turns out to be globally conformally related to the flat metric; the manifold is therefore $M\cong{\mb R}^2$. 

The cubic integral
\beq
S_1=(u(t)\,P_t)^3+\mu(t)(u^2(t)-a)(u(t)\,P_t)\,P_y^2-ay\,P_y^3+2y\,H\,P_y
\eeq
remains hence globally defined, and the same holds true for $S_2$.

- For $a=0$ the function $\dst \mu=1+\frac{1}{u^3}$ is no longer even, so we must consider that $u\in{\mb R}$ 
and the scalar curvature
\[
R_G=2u^6\,\frac{(2-u^3)}{(1+u^3)^3}
\]
is not defined for $u=-1\ $; there is thus no obtainable manifold structure.

- For $a<0$ we set $\rho=|a|$ and we must take $u>0$; we then define the new coordinate
\[
t=u\left(1+\frac{1}{\rho\sqrt{\rho+u^2}}\right),
\qq\qq   
u \in (0,+\nf)\longmapsto t \in (0,+\nf).
\]
Since $\dst\mu=\frac{dt}{du}$ never vanishes, the inverse function 
$u(t)$ is $C^{\nf}([0,+\nf))$. The metric 
\beq
G=\Om^2\,\frac{dt^2+dy^2}{t^2},
\qq%\qq 
\Om(t)=1+\frac{1}{\rho\sqrt{\rho+u^2}},
\qq 
\rho>0,
\quad 
t>0,
\quad 
y\in{\mb R},
\eeq
is again globally conformally related to the canonical metric on the manifold $M\cong{\mb H}^2$. The proof that 
the cubic integrals are also globally defined is the same as in Proposition~\ref{cas12}.~$\quad\Box$

%--------------------------------------------------------------------------------
\subsubsection{Second case: $\mathbf{\eps=-1}$}
%--------------------------------------------------------------------------------

The metric and the Hamiltonian are now given by
\beq
G=\frac{1}{u^2}\Big(\mu^2\,du^2+dy^2\Big),
\qq%\qq 
H=\frac{u^2}2\left(\frac{P_u^2}{\mu^2}+P_y^2\right),
\qq%\qq 
a-u^2>0,
\quad 
y\in\mb{R},
\eeq
where
\[
\mu=1+\frac{C}{(a-u^2)^{3/2}}.
\]
The scalar curvature reads thus
\beq\label{curv3}
R_G=-\frac{2}{\mu^3}\left(1+C\,\frac{(2u^2+a)}{(a-u^2)^{5/2}}\right).
\eeq

The cubic integral $S_1$ is the same as in (\ref{S1cas2}) while
\beq
S_2=y\,S_1+\half\left(a(u^2+y^2)+\frac{2Cu^2}{\sqrt{a-u^2}}+\frac{C^2}{a-u^2}\right)P_y^3
-\left(u^2+y^2+\frac{2C}{\sqrt{a-u^2}}\right)H\,P_y
\eeq
is merely obtained by the substitution $C \to -C$.

%Let us prove:

\begin{nth}Either for $C=-1$ and $\ 0<a<1$ or for $C=+1$ the super\-integrable system $(H,P_y,S_1)$ is globally defined on the manifold $M\cong{\mb H}^2$.
\end{nth}

\nin{\bf Proof:} We must have $a>0$ to ensure $u\in (0,\sqrt{a})$. 

- For $C=-1$ the scalar curvature
is singular when $\mu$ vanishes. This happens for $u_0=\sqrt{a-1}$ and $a\geq 1$; in this case there exists no manifold structure. However for $0<a<1$ the function $\mu$ never vanishes, so we can define
\[
t=-u\left(1-\frac{1}{a\sqrt{a-u^2}}\right),
\qq\qq 
u \in (0,\sqrt{a}) \longmapsto t \in (0,+\nf),
\]
and the inverse function $u(t)$ is in $C^{\nf}([0,+\nf))$; this leads to the metric
\beq
G=\Om^2\,G(\mb{H}^2,{\rm can}),
\qq\qq 
\Om(t)=-1+\frac{1}{a\sqrt{a-u^2(t)}},
\qq
0<a<1,
\eeq
where the conformal factor never vanishes; hence, the manifold is again $M\cong{\mb H}^2$. The proof that 
the cubic integrals are also globally defined is the same as in Proposition \ref{cas12}.

- For $C=+1$ the function
\[
\mu=1+\frac{1}{(a-u^2)^{3/2}}
\]
never vanishes, implying that the curvature is defined everywhere for $u\in (0,\sqrt{a})$. If we define 
\[
t=u\left(1+\frac{1}{a\sqrt{a-u^2}}\right),
\qq\qq 
u \in (0,\sqrt{a}) \longmapsto t \in (0,+\nf),
\]
the metric retains the form
\beq
G=\Om^2\,G(\mb{H}^2,{\rm can}),
\qq\qq 
\Om=1+\frac{1}{a\sqrt{a-u^2(t)}},
\qq 
a>0,
\eeq
where the conformal factor, $\Omega$, never vanishes; hence, the manifold is again ${M\cong\mb H}^2$. At last, the proof that 
the cubic integrals $S_1$ and $S_2$ are also globally defined is the same as in Proposition~\ref{cas12}.~$\quad\Box$

%%%%%%%%%%%%%%%%%%%%%%%%%%%%%%%%%%%%%%%%%%%%%%%%%%%%%%%%%%%%%%%%%%%%%%%%%%%%%%%%%
%%%%%%%%%%%%%%%%%%%%%%%%%%%%%%%%%%%%%%%%%%%%%%%%%%%%%%%%%%%%%%%%%%%%%%%%%%%%%%%%%
\section{Conclusion}\label{conclusionSection}
%%%%%%%%%%%%%%%%%%%%%%%%%%%%%%%%%%%%%%%%%%%%%%%%%%%%%%%%%%%%%%%%%%%%%%%%%%%%%%%%%
%%%%%%%%%%%%%%%%%%%%%%%%%%%%%%%%%%%%%%%%%%%%%%%%%%%%%%%%%%%%%%%%%%%%%%%%%%%%%%%%%

We have completed the work initiated by Matveev and Shevchishin in \cite{ms} by providing the explicit 
form of their metrics in local coordinates. This allowed us to determine systematically
all the cases in which their superintegrable systems can be hosted by a simply-connected, 
two-dimensional smooth manifold $M$. Let us emphasize that we have achieved, via Theorem \ref{Thm1} and Theorem \ref{Thm2}, the classification of all these metrics on closed, simply-connected, surfaces, namely on $M\cong{\mb S}^2$. 
% Of course, we do not rule out the possibility of metrics on non simply-connected surfaces, e.g., 
% on the torus ${\mb T}^2$, but those happen to be very difficult to determine explicitly; 
% they will certainly deserve further study.

{As pointed out in \cite{ms} superintegrable systems on a closed manifold should lead to Zoll metrics \cite{Be}, i.e., to metrics whose geodesics are all closed and of the same length. Using the explicit formulas obtained here for the metrics it has been proved by a direct analysis in \cite{Va3} that all the metrics defined on $S^2$ that we have obtained here are indeed Zoll metrics. Generalizing this analysis to closed  orbifolds gives either Tannery or Zoll metrics.}

Another obvious line of research would be the generalization of these results to the case of observables of fourth or even higher degree, as well as the challenging problem of their 
quantization. {An interesting approach could be to use a well and  uniquely defined quantization procedure, in our case the conformally-equivariant quantization 
\cite{dlo}. The latter, from its very definition and construction, could be perfectly fitted to deal with integrable systems on Riemann surfaces.} 
%It will be interesting to observe in the future what kind of progress can be made in all of these directions.

%%%%%%%%%%%%%%%%%%%%%%%%%%%%%%%%%%%%%%%%%%%%%%%%%%%%%%%%%%%%%%%%%%%%%%%%%%%%%%%%%
%%%%%%%%%%%%%%%%%%%%%%%%%%%%%%%%%%%%%%%%%%%%%%%%%%%%%%%%%%%%%%%%%%%%%%%%%%%%%%%%%
\section{Appendix: the hyperbolic plane}
%%%%%%%%%%%%%%%%%%%%%%%%%%%%%%%%%%%%%%%%%%%%%%%%%%%%%%%%%%%%%%%%%%%%%%%%%%%%%%%%%
%%%%%%%%%%%%%%%%%%%%%%%%%%%%%%%%%%%%%%%%%%%%%%%%%%%%%%%%%%%%%%%%%%%%%%%%%%%%%%%%%

Let us recall that the hyperbolic plane  
\beq
{\mb H}^2=\{(x^1,\,x^2,\,x^3)\in{\mb R}^3\ |\ (x^1)^2+(x^2)^2-(x^3)^2=-1,\ x^3>0\}
\eeq
may be embedded in $\mb{R}^{2,1}$ as follows
%according to the formulas
\beq
x^1=\frac{y}{t},
\qq 
x^2=\frac{1}{2t}(t^2+y^2-1),
\qq 
x^3=\frac{1}{2t}(t^2+y^2+1).
\eeq
This choice of coordinates leads to the induced metric
\beq
G(\mb{H}^2,{\rm can})=\frac{dt^2+dy^2}{t^2},
\qq\qq\qq 
t>0,
\quad 
y\in{\mb R}.
\eeq
The generators on $T^*({\mb H}^2)$ of the group of isometries of ${\mb H}^2$  given by 
\beq\label{iso}
\barr{l}\dst M_1=x^2\,P_3+x^3\,P_2=t\,P_t+y\,P_y, \\[3mm]\dst 
M_2=x^3\,P_1+x^1\,P_3=-ty\,P_t+\frac{(1+t^2-y^2)}{2}\,P_y, \\[3mm]\dst 
M_3=x^1\,P_2-x^2\,P_1=+ty\,P_t+\frac{(1-t^2+y^2)}{2}\,P_y, \earr
\eeq
are globally defined and generate, with respect to the Poisson bracket, the Lie algebra $\mathrm{sl}(2,{\mb R})$, namely
\[
\{M_1,M_2\}=-M_3,
\qq\quad 
\{M_2,M_3\}=M_1,
\qq\quad 
\{M_3,M_1\}=M_2.
\]

%%%%%%%%%%%%%%%%%%%%%%%%%%%%%%%%%%%%%%%%%%%%%%%%%%%%%%%%%%%%%%%%%%%%%%%%%%%%%%%%%
%%%%%%%%%%%%%%%%%%%%%%%%%%%%%%%%%%%%%%%%%%%%%%%%%%%%%%%%%%%%%%%%%%%%%%%%%%%%%%%%%

\end{document}